\documentclass[conference]{IEEEtran}\IEEEoverridecommandlockouts
\usepackage[dvips]{epsfig}
\usepackage[dvips]{graphicx}
\usepackage[dvips]{graphicx,color}
\usepackage{boxedminipage}
\usepackage{color}
\usepackage{amsfonts}
\usepackage{amssymb}
\usepackage{amsmath}
\usepackage{graphics,amsmath,amsfonts,amssymb,epsfig,latexsym,amsfonts,graphicx,amssymb}
\usepackage{eqparbox}

\makeatletter
\def\ps@headings{%
\def\@oddhead{\mbox{}\scriptsize\rightmark \hfil \thepage}%
\def\@evenhead{\scriptsize\thepage \hfil \leftmark\mbox{}}%
\def\@oddfoot{}%
\def\@evenfoot{}}
\makeatother 

\newcommand{\bfbeta}{{\mbox{\boldmath $\beta$}}}

\newcommand{\bfPhi}{{\mbox{\boldmath $\Phi$}}}

\def\BibTeX{{\rm B\kern-.05em{\sc i\kern-.025em b}\kern-.08em
    T\kern-.1667em\lower.7ex\hbox{E}\kern-.125emX}}

\begin{document}
\vspace{-0.5cm}
\title{Joint Distributed Access Point Selection and Power Allocation in Cognitive Radio Networks}
\author{\IEEEauthorblockN{Mingyi Hong, Alfredo Garcia and Jorge Barrera}
\IEEEauthorblockA{ Department of Systems and Information
Engineering, University of Virginia, VA, 22903}\thanks{This work has
been supported in part by the National Science Foundation under
Award CCF-1017982 and IIP-0646008, and through the Wireless Internet
Center for Advanced Technology (WICAT) at University of Virginia.} }
\maketitle

\begin{abstract}
 Spectrum management has been identified as a crucial
step towards enabling the technology of the cognitive radio network
(CRN). Most of the current works dealing with spectrum management in
the CRN focus on a single task of the problem, e.g., spectrum
sensing, spectrum decision, spectrum sharing or spectrum mobility.
In this work, we argue that for certain network configurations,
jointly performing several tasks of the spectrum management improves
the spectrum efficiency. Specifically, we study the uplink resource
management problem in a CRN where there exist multiple cognitive
users (CUs) and access points (APs), with each AP operates on a set
of non-overlapping channels. The CUs, in order to maximize their
uplink transmission rates, have to associate to a suitable AP
(spectrum decision), and to share the channels belong to this AP
with other CUs (spectrum sharing). These tasks are clearly
interdependent, and the problem of how they should be carried out
efficiently and distributedly is still open in the literature.

In this work we formulate this joint spectrum decision and spectrum
sharing problem into a non-cooperative game, in which the feasible
strategy of a player contains a discrete variable and a continuous
vector. The structure of the game is hence very different from most
non-cooperative spectrum management game proposed in the literature.
We provide characterization of the Nash Equilibrium (NE) of this
game, and present a set of novel algorithms that allow the CUs to
distributively and efficiently select the suitable AP and share the
channels with other CUs. Finally, we study the properties of the
proposed algorithms as well as their performance via extensive
simulations.

\end{abstract}
\section{Introduction}
\subsection{Motivation and Related Work}\label{subMotivation}
The problem of distributed spectrum management in the context of CRN
has been under intensive research recently. As pointed out by the
authors of \cite{akyildiz08}, the spectrum management needs to
address four main tasks: 1) {\it spectrum sensing}, techniques that
ensure CUs to find the unused spectrum for communication; 2) {\it
spectrum decision}, protocols that enable the CUs to decide on the
best set of channels; 3) {\it spectrum sharing}, schemes that allow
different CUs to share the same set of channels; 4) {\it spectrum
mobility}, rules that require the CUs to leave the channel if
licensed users are detected. Many efforts have been devoted to
providing solutions to the individual tasks listed above. However,
as we will see in this paper, in some CRN scenarios, several of the
above tasks become interdependent, and the CUs have to perform these
tasks {\it jointly} to achieve best performance. In this work, we
propose to provide solutions for the joint spectrum decision and
spectrum sharing problems in a multi-channel multi-user CRN.

We focus on investigating an important CRN scenario where such joint
spectrum decision and spectrum sharing is desirable. Consider a
network with multiple CUs and APs, where the APs operate on
different sets of channels, and the CUs need to connect to one of
the APs for communication. The CUs can use {\it multiple channels}
belong to the associated AP concurrently for transmission, but
different CUs interfere with each other if they use the same
channel. This network is a generalization of the single AP network
considered in many previous literature, e.g., \cite{lai08} and
\cite{islam08}. In the considered network, the CUs face the spectrum
decision problem when they select the AP, and they face the spectrum
sharing problem when they try to dynamically allocate their
communication power across the channels belong to the selected AP.
Clearly, these two problems are strongly interdependent, as on the
one hand a particular CU has to select an AP before it can share the
spectrum that belongs to this AP with all the other CUs associated
with it; on the other hand, after sharing the spectrum, an
individual CU may have the incentive to switch to a different AP if
it perceives that such action will increase its communication rate.
A poor spectrum decision and spectrum sharing scheme will not only
lead to unsatisfactory performance for individual CUs, but also
result in an unstable system in which CUs are constantly unsatisfied
with their current communication rates and consequently changing
their AP associations and power allocation indefinitely.

%

A similar problem related to the joint cell selection/base station
(BS) association and power control has been addressed in
infrastructure-based cellular networks. \cite{hanly95} and
\cite{yates95b} are early works trying to tackle this problem in an
uplink spread spectrum cellular network. The objectives are to let
the users find a best site selection and power allocation tuple such
that all users' target signal to interference ratio (SIR) are met,
and all users' transmission power is minimized. The authors of
\cite{apcan06} and \cite{saraydar01} cast a similar problem (with an
objective to maximize individual power efficiency or minimizing
individual cost) into game theoretical frameworks, and propose
algorithms to find the Nash Equilibrium (NE) of the proposed games.
One of the most important differences between our work and the above
cited works is that the power allocation problems in these works are
essentially {\it scalar value} optimization problem: each user only
needs to decide on its {\it power level} once a BS is selected,
while in our work, individual power allocation is a {\it vector
optimization} problem as the CUs have the flexibility to use {\it
all} the channels belong to a particular AP concurrently. This
fundamental difference makes the considered problem more complex,
hence the analytical frameworks provided by the above cited works
are not suitable for our problem. It can also be argued that the
problem under consideration is also in many aspects more complicated
than the traditional AP association problems arise in the 802.11
WLAN network (for example, \cite{kauffmann07}, \cite{bejerano07} and
\cite{kumar05} and the reference therein). Typically, AP association
is aiming to optimize different system performance metrics
(throughput, fairness, etc), and only simple individual throughput
estimates within each AP are used to update the current association
profile. Indeed, in 802.11 WLAN network, the throughput of an
individual AP with fixed number of users and fixed physical bit rate
can be approximated using simple analytical formulae \cite{kumar07},
and this result has greatly simplified the analysis of many work
dealing with dynamic AP association in WLAN, e.g., \cite{kumar05}
and \cite{shakkottai07}.

We also note here that the problem of how to dynamically perform the
task of both spectrum decision and spectrum sharing may arise in
other important CRN configurations as well. Most of the current
works addressing the spectrum management problem in multi-channel
multi-user CRN focus only on the spectrum sharing part of the
problem. For example, in \cite{wang08}, \cite{wu09}, a set of
iterative water-filling (IWF) \footnote{IWF is originally proposed
in \cite{yu02a} for DSL network, and subsequently applied to
wireless networks. See e.g., \cite{scutari08a} and the references
therein.} based algorithms are proposed to find a distributed
solution of power allocation in multi-channel, multi-user CRN with
interference channels. One important assumption underlying these
works is that the CUs are able to use {\it all the channels
simultaneously}. However, this assumption might not be valid in the
situation where the available spectrum is fragmented due to licensed
user activities and where the CUs are equipped with 1-agile radio
which can only use a single set of {\it continuously aligned}
channels at a time \footnote{see \cite{cao10} for detailed
discussion for the possibility that this scenario might rise in
actual CRN implementations}. In this scenario, the CUs need to select
the set of channels to use, and decide on the allocation of the
transmission power to the selected set of channels, i.e., the CUs
are required to perform the task of joint spectrum decision and
spectrum sharing. Although the problem of how to optimally perform
such task has never been addressed in literature before, it is our
belief that our work can also serve to shed some lights on providing
solutions to it, as the network configuration considered in our work
is sufficiently similar to the configuration mentioned above.



\subsection{Contributions and Organization of This Work}
To the best of our knowledge, this is the first work that propose
distributed algorithms to deal with joint AP selection and power
allocation problem in a multi-channel multi-AP CRN. We cast the
problem into a non-cooperative game framework, in which each CU's
objective is to maximize its own transmission rate, and its strategy
space is the union of a {\it discrete set} (the set of possible APs)
and a {\it multi-dimensional continuous set} (the set of feasible
power vectors). Although non-cooperative game theory has recently
been extensively applied to solve the resource allocation problem in
CRN (e.g., \cite{wang08} and \cite{wu09} and the reference therein),
our formulation is considerably different and more involved because
of such ``hybrid" nature of the strategy space of the game. We
analyze in detail the equilibrium solution of the game, and develop
an algorithms with provable convergence guarantees that enables the
CUs to distributedly compute the equilibrium solution. Finally, we
suggest various extensions of our original algorithm based on
practical considerations.

We organize the paper as follows. In section \ref{secSystemModel},
we present the system model and formulate the problem into a
non-cooperative game. In section \ref{secNE}, we analyze the
properties of the NE. In section \ref{secAlgorithm}, we provide our
main algorithm and its convergence properties. In section
\ref{secExtension}, we provide extensions of the JASPA algorithm. We
present simulation results in section \ref{secSimluation} and
conclude the paper in section \ref{secConclusion}.

%
\vspace{-0.1cm}
\section{Problem Statement and System Model}\label{secSystemModel}
\subsection{Considered Network and Some Assumptions}
We consider the following cognitive network configuration. Suppose
there are a set $\{1,\cdots,N\}\triangleq\mathcal{N}$ CUs,
$\{1,\cdots,K\}\triangleq\mathcal{K}$ channels and
$\{1,\cdots,W\}\triangleq\mathcal{W}$ APs in the network. Each AP
$w\in\mathcal{W}$ is assigned with a subset of channels
$\mathcal{K}_w\subseteq\mathcal{K}$. We focus on the uplink scenario
where each CU wants to connect to one of the APs for transmission.

The followings are our main assumptions of the network.\\
{\bf A-1)} Each CU $i$ is able to associate to all the APs, and each
AP covers entire area of the network.\\
{\bf A-2)} The APs covering the same area operate on non-overlapping
portions of the available spectrum.\\
{\bf A-3)} The set of available spectrum can be used exclusively by
the CRN, for a relative long period of time.\\
{\bf A-4)} Each CU can associate to a single AP at a time; it can
concurrently use all the channels of the associated AP.

Assumption A-1) is made merely for ease of presentation, and our
work can be extended to the scenarios where different APs cover
different areas of the network, and where the CUs can only connect
to the subset of APs that cover them.

Assumption A-2) is commonly used when considering AP association
problems in WLAN (for example, in \cite{shakkottai07}), and it is
made to mitigate interference between neighboring APs. It can be
achieved either by 1) the APs agree offline the partition of the
spectrum, or 2) the APs jointly run a distributed online spectrum
assignment algorithm similar to the ones proposed in
\cite{kauffmann07} to determine the best spectrum assignment.
Assumption A-1) and A-2) imply that
$\mathcal{K}_q\bigcap\mathcal{K}_w=\emptyset,~\forall~q\ne w,
q,w\in\mathcal{W}$.


Assumption A-3) can be achieved either under the spectrum property
right model in which the licensed networks sell or lease the
spectrum to the cognitive network for a period of time for exclusive
use, or under the situation that the cognitive network exploits
relative static spectrum white spaces unused by local TV broadcast
\cite{zhao07}.

\subsection{System Model and A Non-Cooperative Game Formulation}
Let $\{|h_{i,w}(k)|^2\}_{k\in\mathcal{K}_w}$ be the set of power
gains from CU $i$ to AP $w$ on all its channels; Let
$\{n_w(k)\}_{k\in\mathcal{K}_w}$ be the set environmental noise
powers on all channels for AP $w$; Let the $N\times 1$ vector
$\mathbf{a}$ denote the {\it association profile} in the network,
with its $i^{th}$ element $\mathbf{a}(i)=w$ indicating that CU $i$
is associated to AP $w$. Each CU $i$ is able to obtain {\it its own}
channel gains to all the APs,
$\{|h_{i,w}(k)|^2\}_{k\in\mathcal{K}_w, w\in\mathcal{W}}$, via
feedback from the APs, but it {\it does not} need to have the
knowledge of other CUs' channel gains in the network.

Let ${p}_{i,w}(k)$ represent the amount of power CU $i$ transmits on
channel $k$ when it is associated with AP $w$; Let
$\mathbf{p}_{i,w}=\left\{p_{i,w}(k)\right\}_{k\in\mathcal{K}_w}$ be
the {\it power profile} of CU $i$ when it is associated with AP $w$;
let $\mathbf{p}_{-i,w}$ be the joint power profiles of all the CUs
other than $i$ that is associated with AP $w$:
$\mathbf{p}_{-i,w}\triangleq\{\mathbf{p}_{j,w}\}_{j:j\ne i,
\mathbf{a}(j)=w}$. By construction, for all $w\in\mathcal{W}$, if
$w\ne\mathbf{a}(i)$, then $\mathbf{p}_{i,w}= \mathbf{0}$. The power
profiles of the CUs must also satisfy the following two constraints:
1) Total power constraints: $\sum_{k\in\mathcal{K}_w}p_{i,w}(k)\le
\bar{p}_i,~\forall~i$, where $\bar{p}_i$ is the power limit for CU
$i$; 2) Positivity constraints: $p_{i,w}(k)\ge 0~\forall~i,~k$. As
such, each CU's feasible power allocation when it is associated with
AP $w$ can be expressed as:{
\begin{align}
\vspace{-0.15cm}\mathcal{F}_{i,w}\triangleq\Big\{\mathbf{p}_{i,w}:\hspace{-0.1cm}\sum_{k\in\mathcal{K}_w}p_{i,w}(k)\le
\bar{p}_i,~p_{i,w}(k)\ge
0,~\forall~k\in\mathcal{K}_w\Big\}.\nonumber
\end{align}}
Assume there is no interference cancelation performed at the AP, and
the interference caused by other CUs are treated as noises by each
CU. As mentioned in \cite{lai08}, this assumption is reasonable
considering the lack of coordination among the CUs. It further
allows for the implementation of low-complexity single-user decoders
on the AP. Given this assumption, for a fixed AP association and
power allocation configuration, CU $i$'s uplink transmission rate
(when it is associated with AP $w$) can be expressed as
follows:{\vspace{-0.15cm}
\begin{align}
&\hspace{-0.25cm}R_i(\mathbf{p}_{i,w},\mathbf{p}_{-i,w};
w)\nonumber\\
&\hspace{-0.25cm}=\hspace{-0.15cm}\sum_{k\in\mathcal{K}_w}\hspace{-0.12cm}
\log\hspace{-0.08cm}\Big(\hspace{-0.05cm}1\hspace{-0.08cm}+\hspace{-0.08cm}\frac{|h_{i,w}(k)|^2p_{i,w}(k)}{n_w(k)+\sum_{j:\mathbf{a}(j)=w, j\ne i}|h_{j,w}(k)|^2p_{j,w}(k)}\Big)\label{eqTransmissionRate}\\
&\hspace{-0.25cm}=\hspace{-0.12cm}\sum_{k\in\mathcal{K}_w}\hspace{-0.12cm}
\log\hspace{-0.08cm}\Big(\hspace{-0.05cm}1+\frac{|h_{i,w}(k)|^2p_{i,w}(k)}{n_w(k)+I_{i}(k)}\Big)\triangleq
R_i(\mathbf{p}_{i,w},\mathbf{I}_{i,w};w)\label{eqRAlternativeDefinition}
\end{align}}
where ${I}_i(k)$ denotes the aggregated received transmission power
level on channel $k$
 except CU $i$, i.e.,
\begin{align}
I_i(k)\triangleq\hspace{-0.5cm}\sum_{j:\mathbf{a}(j)=w, j\ne
i}\hspace{-0.4cm}|h_{j,w}(k)|^2p_{j,w}(k),~~
\mathbf{I}_{i,w}\triangleq\left\{I_i(k)\right\}_{k\in\mathcal{K}_w}\label{eqI}.
\end{align}
Clearly, if $w=\mathbf{a}(i)$, then
$\left\{I_i(k)\right\}_{k\in\mathcal{K}_w}$ can be viewed as the set
of aggregated interference {\it currently experienced} by CU $i$; if
$w\ne \mathbf{a}(i)$, $\left\{I_i(k)\right\}_{k\in\mathcal{K}_w}$
can also be viewed as the set of aggregated interference for CU $i$
{\it if it were to switch to AP w}.

We see that \eqref{eqTransmissionRate} and
\eqref{eqRAlternativeDefinition} are equivalent definitions of the
CU $i$'s transmission rate. We will use either definition in the
following paragraph depending on the context.

We model each CU $i$ as selfish agent, and its objective is to find
strategy $(w^*,\mathbf{p}^{*}_{i,w^*})$ that maximizes its
transmission rate:\vspace{-0.1cm}
\begin{align}
\hspace{-0.1cm}\big(w^*,\mathbf{p}^{*}_{i,w^*}\big)\in
\arg\max_{w\in\mathcal{W}}\max_{p_{i,w}\in\mathcal{F}_{i,w}}R_i(\mathbf{p}_{i,w},\mathbf{p}_{-i,w};
w).
\end{align}
We are now ready to define a non-cooperative game $\mathcal{G}$:
\begin{align}
\mathcal{G}\triangleq\left\{\mathcal{N},\{{\chi}_i\}_{i\in\mathcal{N}},
\{R_i\}_{i\in\mathcal{N}}\right\}\label{eqGame}
\end{align}
where the CUs $i\in\mathcal{N}$ are the players in the game; each
CU's strategy space can be expressed as $\chi_i\triangleq
\bigcup_{w\in\mathcal{W}}\left\{w, \mathcal{F}_{i,w}\right\}$;
each CU's utility function is its transmission rate
$R_i(\mathbf{p}_{i,w},\mathbf{p}_{-i,w}; w)$ as defined in
\eqref{eqTransmissionRate}. We emphasize that each feasible strategy
of a player in the game $\mathcal{G}$ contains a discrete variable
and a continuous vector, which makes the game $\mathcal{G}$ unique
to (and thus more complicated than) most of the games considered in
the context of network resource allocation. We refer to the strategy
space $\{\chi_i\}$ of this game as {\it hybrid strategy space}.

The NE of this game is defined as the tuple $\left\{\mathbf{a}^*(i),
\mathbf{p}^*_{i,{\mathbf{a}}^*(i)}\right\}_{i\in\mathcal{N}}$ such
that for all $i\in\mathcal{N}$ the following set of equations are
satisfied:\vspace{-0.1cm}
\begin{align}
\hspace{-0.1cm}\Big(\mathbf{a}^*(i),\mathbf{p}^{*}_{i,\mathbf{a}^*(i)}\Big)\in
\arg\max_{w\in\mathcal{W}}\max_{p_{i,w}\in\mathcal{F}_{i,w}}R_i(\mathbf{p}_{i,w},\mathbf{p}^{*}_{-i,w};
w)\label{eqNE}
\end{align}
or equivalently,\vspace{-0.1cm}
$~\forall~i\in\mathcal{N},~w\in\mathcal{W},~\mathbf{p}_{i,w}\in\mathcal{F}_{i,w}$,
\begin{align}
R_i(\mathbf{p}_{i,\mathbf{a}^*(i)}^*,\mathbf{p}^{*}_{-i,\mathbf{a}^*(i)};
\mathbf{a}^*(i))\ge R_i(\mathbf{p}_{i,w},\mathbf{p}^{*}_{-i,w};
w)\nonumber.
\end{align}
We call the equilibrium profile $\mathbf{a}^*$ a {\it NE association
profile}, and
$\mathbf{p}^*_{\mathbf{a}^*}\triangleq\big\{\mathbf{p}^*_{i,{\mathbf{a}}^*(i)}\big\}_{i\in\mathcal{N}}$
a {\it NE power allocation profile}. In order to avoid duplicated
definitions, we call the tuple $\left(\mathbf{a}^*,
\mathbf{p}^*_{\mathbf{a}^*}\right)$ a {\it joint equilibrium
profile} (JEP) of the game $\mathcal{G}$ (instead of a NE). It is
clear from either of the above definitions that in a JEP, the system
is stable in the sense that no CU has the incentive to deviate from
either its AP association or its power allocation.

\section{Properties of the JEP}\label{secNE}
In this section, we introduce the notion of the potential function,
and characterize its relationship with the JEP. We then prove that
the JEP always exists for the game $\mathcal{G}$. The proof of Lemma
\ref{lemmaPotentialMaximization} can be found in
\cite{hong10m_japsa}, which is an extended version of this paper.

Let us consider a simpler problem in which the association vector
$\mathbf{a}$ is {\it predetermined} and {\it fixed}. In this case,
the CUs do not need to choose their AP associations, thus the
problem of finding the JEP defined in \eqref{eqNE} reduces to the
one of finding the NE power allocation profile
$\mathbf{p}^*_{\mathbf{a}}$ that satisfies:
\begin{align}
\mathbf{p}^*_{i,\mathbf{a}(i)}\in\arg\max_{\mathbf{p}_i\in\mathcal{F}_{i,\mathbf{a}(i)}}R_i(\mathbf{p}_i,
\mathbf{p}^*_{-i,\mathbf{a}(i)};\mathbf{a}(i)).\label{eqNEPowerProfile}
\end{align}
For a specific AP $w$, denote the set of CUs associated with it to
be $\mathcal{N}_w$: $\mathcal{N}_w\triangleq\{i:\mathbf{a}(i)=w\}$.
We use
$\mathbf{p}_{w}\triangleq\left\{\mathbf{p}_{i,w}\right\}_{i\in\mathcal{N}_w}$
to denote the long vector containing the power profiles of all CUs
associated with AP $w$. When $\mathbf{a}$ is fixed, the activity of
the set of CUs $\mathcal{N}_w$, $w\in\mathcal{W}$ does not affect
the activity of the set of CUs $\mathcal{N}_q$, $q\in\mathcal{W},
q\ne w$, because AP $w,q$ operate on different sets of channels:
$\mathcal{K}_w\bigcap \mathcal{K}_q=\emptyset$. Consequently, the
original game $\mathcal{G}$ introduced in \eqref{eqGame} can be
decomposed into $W$ independent small games
$\left\{\mathcal{G}^{\mathbf{a}}_w\right\}_{w=1}^{W}$, with each
small game defined as:
\begin{align}
\mathcal{G}^{\mathbf{a}}_w\triangleq\left\{\mathcal{N}_w,\{{\mathcal{F}_{i,w}}\}_{i\in\mathcal{N}_w},
\{R_i\}_{i\in\mathcal{N}_w}\right\}.\label{eqSubGame}
\end{align}
From the standard theory regarding to the existence of the NE, it is
straightforward to see that there exists at least one NE power
allocation $\mathbf{p}_w^*(\mathbf{a})$ for each game
$\mathcal{G}^{\mathbf{a}}_w$. In order to further characterize the
NE power profile $\mathbf{p}^*_{w}(\mathbf{a})$ of the game
$\mathcal{G}^{\mathbf{a}}_w$, we first introduce the notion of a
{\it potential function}.
\newtheorem{D1}{Definition}
\begin{D1}
{\it The {\it potential function of the game
$\mathcal{G}_w^{\mathbf{a}}$} under a feasible power profile
$\mathbf{p}_w$ is defined as:\vspace{-0.1cm}
\begin{align}
P_w(\mathbf{p}_w;{\mathbf{a}})=\sum_{k\in\mathcal{K}_w}\log\Big(n_w(k)+\sum_{i\in\mathcal{N}_w}|h_{i,w}(k)|^2
p_{i,w}(k)\Big)\nonumber.\vspace{-0.3cm}
\end{align}
The {\it system potential function} under a specific $\mathbf{a}$
and a feasible $\mathbf{p}$ is defined as the sum of the potential
functions associated to all games
$\{\mathcal{G}_w^{\mathbf{a}}\}_{w\in\mathcal{W}}$: $
P(\mathbf{p};\mathbf{a})=\sum_{w\in\mathcal{W}}P_w(\mathbf{p}_w;\mathbf{a}).
$ }
\end{D1} Clearly, for fixed $\mathbf{a}$,
$P_w(\mathbf{p}_w;{\mathbf{a}})$ is a concave function and hence has
a unique maximum point. Define
$\mathcal{F}^{\mathbf{a}}_{w}\triangleq
\prod_{i\in\mathcal{N}_w}\mathcal{F}_{i,w}$ as the joint feasible
set for the CUs that are associated with AP $w$ under the
association profile $\mathbf{a}$, and let
$\mathcal{F}^{\mathbf{a}}\triangleq
\prod_{w\in\mathcal{W}}\mathcal{F}^{\mathbf{a}}_{w}$. Let
$\mathcal{E}_w(\mathbf{a})$ denote the set of all NE power profiles
for the game $\mathcal{G}^{\mathbf{a}}_w$, then
$\mathcal{E}(\mathbf{a})\triangleq
\prod_{w\in\mathcal{W}}\mathcal{E}_w(\mathbf{a})$ is the set of all
NE power profiles for the game $\mathcal{G}$ under fixed association
profile $\mathbf{a}$. Let
$\mathbf{p}^*_w(\mathbf{a})\in\mathcal{E}_w(\mathbf{a})$ and
$\mathbf{p}^*(\mathbf{a})\in\mathcal{E}(\mathbf{a})$, then we have
the following lemma regarding to the relationship between
$\mathbf{p}^*_w(\mathbf{a})$, $\mathbf{p}^*(\mathbf{a})$ and the
potential functions.
\newtheorem{L1}{Lemma}
\begin{L1}\label{lemmaPotentialMaximization}
{\it For fixed $\mathbf{a}$, a feasible
$\mathbf{p}_{w}\in\mathcal{F}^\mathbf{a}_{w}$ maximizes the
potential function $P_w(\mathbf{p}_w;{\mathbf{a}})$ if and only if
it is in the set $\mathcal{E}_w(\mathbf{a})$.
We define the {\it unique maximum value} of the potential function
as the Equilibrium Potential (EP) for AP $w$ under association
profile $\mathbf{a}$: $ \bar{P}_w(\mathbf{a})\triangleq
\max_{\mathbf{p}_w\in\mathcal{F}^{\mathbf{a}}_w}P_w(\mathbf{p}_w;\mathbf{a})$.

For a fixed $\mathbf{a}$, a feasible
$\mathbf{p}\in\mathcal{F}^{\mathbf{a}}$ that maximizes the system
potential function $P(\mathbf{p};\mathbf{a})$ iff it is in the set
$\mathcal{E}(\mathbf{a})$. Similarly as above, we refer to the
unique maximum value of the system potential function as the System
Equilibrium Potential (SEP) under $\mathbf{a}$, and denote it by
$\bar{P}(\mathbf{a})$: $
\bar{P}(\mathbf{a})\triangleq\sum_{w\in\mathcal{W}}\bar{P}_w(\mathbf{a}).
$}
\end{L1}


We are now ready to discuss the existence of the JEP of the game
$\mathcal{G}$ as defined in \eqref{eqNE}. We emphasize here that
determining the existence of the JEP (which is a {\it pure} NE) for
the game $\mathcal{G}$ is by no means a trivial proposition. Due to
the hybrid structure of the game $\mathcal{G}$, the standard results
on the existence of pure NE of either continuous or discrete games
can not be applied. 
Consequently, we have to explore the structure of the problem in
proving the existence of JEP for the game $\mathcal{G}$.  We have
the following theorem regarding to the existence of JEP.
\newtheorem{T1}{Theorem}
\begin{T1}\label{theoremExistence}
{\it The game $\mathcal{G}$ always admits a JEP. An association
profile
$\widetilde{\mathbf{a}}\in\arg\max_{\mathbf{a}}\bar{P}(\mathbf{a})$,
along with any one of its corresponding NE power allocation profile
${\mathbf{p}}^*(\widetilde{\mathbf{a}})=
\Big\{\mathbf{p}^*_{i,\widetilde{\mathbf{a}}(i)}\Big\}_{i\in\mathcal{N}}\in\mathcal{E}(\widetilde{\mathbf{a}})$,
constitute a JEP of the game $\mathcal{G}$.}
\end{T1}
\begin{proof}
We prove this theorem by contradiction. Suppose
$\widetilde{\mathbf{a}}$ maximizes the system potential, but
$\widetilde{\mathbf{a}}$ is not a NE association profile. Then there
must exist a CU $i$ who prefers $\widehat{w}\ne \widetilde{w}$.
Define a new association profile $\widehat{\mathbf{a}}$ as:
$\widehat{\mathbf{a}}(j)=\widetilde{\mathbf{a}}(j)$ except for the
$i^{th}$ entry, in which $\widehat{\mathbf{a}}(i)=\widehat{w}$. Let
$\mathbf{p}^*(\widetilde{\mathbf{a}})\in\mathcal{E}(\widetilde{\mathbf{a}})$,
and
$\mathbf{p}^*(\widehat{\mathbf{a}})\in\mathcal{E}(\widehat{\mathbf{a}})$.
The maximum rate CU $i$ can get after switching to $\widehat{w}$
{\it if all other CUs do not change their actions}
is:\vspace{-0.1cm}
\begin{align}
\hspace{-0.1cm}&\widehat{R}_i(\bar{\mathbf{p}}_{i,\widehat{w}},\mathbf{p}^*_{\widehat{w}}(\widetilde{\mathbf{a}});{\widehat{w}})
\hspace{-0.1cm}=\hspace{-0.3cm}\sum_{k\in\mathcal{K}_{\widehat{w}}}
\hspace{-0.1cm}\log\hspace{-0.1cm}\Big(\hspace{-0.05cm}\frac{n_{\widehat{w}}(k)\hspace{-0.1cm}+
\hspace{-0.1cm}I^*_{i}(k)\hspace{-0.1cm}+\hspace{-0.1cm}|h_{i,\widehat{w}}(k)|^2
\bar{p}_{i,\widehat{w}}(k)}{n_{\widehat{w}}(k)\hspace{-0.1cm}+\hspace{-0.1cm}I^*_{i}(k)}\hspace{-0.05cm}\Big)\nonumber\\
\hspace{-0.1cm}&=P_{\widehat{w}}(\bar{\mathbf{p}}_{i,\widehat{w}},\mathbf{p}^*_{\widehat{w}}(\widetilde{\mathbf{a}});{\widehat{\mathbf{a}}})-
P_{\widehat{w}}(\mathbf{p}_{\widehat{w}}^*(\widetilde{\mathbf{a}});{\widetilde{\mathbf{a}}})\label{eqEstimatedRate}
\end{align}
where $I^*_i(k)$ is defined similarly as in \eqref{eqI}, and the
vector $\bar{\mathbf{p}}_{i,\widehat{w}}$ is defined as: $
\bar{\mathbf{p}}_{i,\widehat{w}}=
\arg\max_{\mathbf{p}_i\in\mathcal{F}_{i,\widehat{w}}}
\widehat{R}_i(\mathbf{p}_i,\mathbf{p}^*_{\widehat{w}}(\widetilde{\mathbf{a}});{\widehat{w}}).
$ We can view the rate
$\widehat{R}_i(\bar{\mathbf{p}}_{i,\widehat{w}},\mathbf{p}^*_{\widehat{w}}(\widetilde{\mathbf{a}});{\widehat{w}})$
as CU $i$'s {\it estimate} of the maximum rate it can get if it were
to switch to AP $\widehat{w}$.

Because CU $i$ prefers $\widehat{w}$, from the definition of the JEP
\eqref{eqNE} we see that its current communication rate must be
strictly less than its estimated maximum rate, i.e.:\vspace{-0.15cm}
\begin{align}
{R}_i(\mathbf{p}^{*}_{i,\widetilde{w}}(\widetilde{\mathbf{a}}),\mathbf{p}^{*}_{-i,\widetilde{w}}(\widetilde{\mathbf{a}});{\widetilde{w}})<
\widehat{R}_i(\bar{\mathbf{p}}_{i,\widehat{w}},\mathbf{p}^*_{\widehat{w}}(\widetilde{\mathbf{a}});{\widehat{w}})
\label{eqRtLessThenRBarT+1}
\end{align}
where
$R_i(\mathbf{p}^{*}_{i,\widetilde{w}}(\widetilde{\mathbf{a}}),\mathbf{p}^{*}_{-i,\widetilde{w}}(\widetilde{\mathbf{a}});{\widetilde{w}})$
is the {\it actual} transmission rate for CU $i$ in the association
profile $\widetilde{\mathbf{a}}$:\vspace{-0.25cm}
\begin{align}
&{R}_i({\mathbf{p}}^*_{i,\widetilde{w}}(\widetilde{\mathbf{a}}),\mathbf{p}^*_{-i,\widetilde{w}}(\widetilde{\mathbf{a}});{\widetilde{w}})=\sum_{k\in\mathcal{K}_{\widetilde{w}}}
\log\Big(1+\frac{|h_{i,\widetilde{w}}(k)|^2
{p}^*(k)}{n_{\widetilde{w}}(k)+I^*_{i}(k)}\Big)\nonumber\\
&=P_{\widetilde{w}}({\mathbf{p}}^*_{\widetilde{w}}(\widetilde{\mathbf{a}});{\widetilde{\mathbf{a}}})-
P_{\widetilde{w}}(\mathbf{p}^*_{-i,\widetilde{w}}(\widetilde{\mathbf{a}});{\widetilde{\mathbf{a}}}).\label{eqEquilibriumRate}
\end{align}
Combining \eqref{eqEstimatedRate}, \eqref{eqRtLessThenRBarT+1} and
 \eqref{eqEquilibriumRate} we must have that:
\begin{align}
P_{\widetilde{w}}({\mathbf{p}}^*_{\widetilde{w}}(\widetilde{\mathbf{a}});{\widetilde{\mathbf{a}}})-&
P_{\widetilde{w}}(\mathbf{p}_{-i,\widetilde{w}}^*(\widetilde{\mathbf{a}});{\widetilde{\mathbf{a}}})<\nonumber\\
&P_{\widehat{w}}(\bar{\mathbf{p}}_{i,\widehat{w}},\mathbf{p}^*_{\widehat{w}}(\widetilde{\mathbf{a}});{\widehat{\mathbf{a}}})-
P_{\widehat{w}}(\mathbf{p}^*_{\widehat{w}}(\widetilde{\mathbf{a}});{\widetilde{\mathbf{a}}})\label{eqComparePotential1}.
\end{align}
Notice that the term
$P_{\widetilde{w}}(\mathbf{p}_{-i,\widetilde{w}}^*(\widetilde{\mathbf{a}});{\widetilde{\mathbf{a}}})$
is equivalent to the term
$P_{\widetilde{w}}(\mathbf{p}_{-i,\widetilde{w}}^*(\widetilde{\mathbf{a}});{\widehat{\mathbf{a}}})$,
due to the equivalence of the following sets:
\begin{align}
\{j:j\ne i, \widetilde{\mathbf{a}}(j)=\widetilde{w}\}=\{j:j\ne i,
\widehat{\mathbf{a}}(j)=\widetilde{w}\}\label{eqNumberOfCUSameWHatWStar}.
\end{align}
Recall that Lemma \ref{lemmaPotentialMaximization} says the NE power
allocation profile maximizes the potential function when
$\mathbf{a}$ is fixed:
$\mathbf{p}_{\widetilde{w}}^*(\widehat{\mathbf{a}})\in
\max_{\mathbf{p}_{\widetilde{w}}\in\mathcal{F}^{\widehat{\mathbf{a}}}_{\widetilde{w}}}P_{\widetilde{w}}(\mathbf{p}_{\widetilde{w}};{\widehat{\mathbf{a}}})$.
Observe that the set of CUs associated with AP $\widetilde{w}$ under
profile $\widehat{\mathbf{a}}$ is the same as the set of CUs
associated with AP $\widetilde{\mathbf{a}}$ under profile
$\widetilde{\mathbf{a}}$ excluding CU $i$, we must have
$\mathbf{p}_{-i,\widetilde{w}}^*(\widetilde{\mathbf{a}})\in\mathcal{F}^{\widehat{\mathbf{a}}}_{\widetilde{w}}$.
Consequently, the following is true:\hspace{-0.3cm}
\begin{align}
 P_{\widetilde{w}}({\mathbf{p}}^*_{\widetilde{w}}(\widehat{\mathbf{a}});{\widehat{\mathbf{a}}})& \ge
 P_{\widetilde{w}}(\mathbf{p}^*_{-i,\widetilde{w}}(\widetilde{\mathbf{a}});{\widehat{\mathbf{a}}})\stackrel{(a)}=P_{\widetilde{w}}(\mathbf{p}_{-i,\widetilde{w}}^*(\widetilde{\mathbf{a}});\widetilde{\mathbf{a}})\label{eqPotentialOPT}
\end{align}
 where $(a)$ is from \eqref{eqNumberOfCUSameWHatWStar}. Similarly, we have
 that:
\begin{align}
P_{\widehat{w}}({\mathbf{p}}^*_{\widehat{w}}(\widehat{\mathbf{a}});{\widehat{\mathbf{a}}})\ge
P_{\widehat{w}}(\bar{\mathbf{p}}_{i,\widehat{w}},\mathbf{p}_{\widehat{w}}^*(\widetilde{\mathbf{a}});{\widehat{\mathbf{a}}})
.\label{eqPotentialOPT+1}
\end{align}
Combining \eqref{eqPotentialOPT}, \eqref{eqPotentialOPT+1} and
\eqref{eqComparePotential1}, we have that:
\begin{align}
P_{\widetilde{w}}({\mathbf{p}}_{\widetilde{w}}^*(\widetilde{\mathbf{a}});\widetilde{\mathbf{a}})+&
P_{\widehat{w}}(\mathbf{p}_{\widehat{w}}^*({\widetilde{\mathbf{a}}});
\widetilde{\mathbf{a}})<\nonumber\\
&P_{\widehat{w}}({\mathbf{p}}^*_{\widehat{w}}(\widehat{\mathbf{a}});{\widehat{\mathbf{a}}})+
P_{\widetilde{w}}({\mathbf{p}}^*_{\widetilde{w}}(\widehat{\mathbf{a}});{\widehat{\mathbf{a}}}).
\label{eqTwoTermSumPotential}
\end{align}
Noticing that the equilibrium potentials of all the APs other than
$\widetilde{w}$ and $\widehat{w}$ are the same between the profile
$\widetilde{\mathbf{a}}$ and $\widehat{\mathbf{a}}$, thus adding
them to both sides of \eqref{eqTwoTermSumPotential} we have that:
\begin{align}
\sum_{w\in\mathcal{W}}P_{w}({\mathbf{p}}^*_{
{w}}(\widetilde{\mathbf{a}});{\widetilde{\mathbf{a}}})<
\sum_{w\in\mathcal{W}}P_w({\mathbf{p}}^*_w(\widehat{\mathbf{a}});{\widehat{\mathbf{a}}})\label{eqSumPotentialInequality}
\end{align}
which is equivalent to:
$\bar{P}(\widetilde{\mathbf{a}})<\bar{P}(\widehat{\mathbf{a}}). $
This is a contradiction to the assumption that
$\bar{P}(\widetilde{\mathbf{a}})$ is the maximum system potential.
We conclude that $\widetilde{\mathbf{a}}$ must be a NE association
profile. Clearly, $\mathbf{p}^*(\widetilde{a})$ is a NE power
allocation profile. Consequently, we have that
$\left(\widetilde{\mathbf{a}}, \mathbf{p}^*(\widetilde{\mathbf{a}})
\right)$ is a JEP.
\end{proof}

\section{The JASPA Algorithm and Its Convergence}\label{secAlgorithm}
In this section, we first introduce an algorithm that assigns the
CUs to their closest AP. This algorithm, although simple and
inefficient, offers valuable insights upon which we build our first
algorithm, called the Joint Access point Selection and Power
Allocation (JASPA) algorithm, in subsection \ref{subJASPA}.

\subsection{The Closest AP Association Algorithm}\label{subNearestAP}
Consider a fixed AP association profile $\mathbf{a}$ in which each
CU is assigned to its closest AP. The ``closeness" from a CU to the
APs can be measured either by the physical distance, or by the
strength of pilot signals received by the CU from the APs. Assuming
that each CU has a single closest AP (ties are randomly broken),
then the AP association profile is fixed and the computation of JEP
reduces to
the problem of finding the NE power allocation profile. 
 Clearly, this scheme {\it separates} the process of spectrum
decision and spectrum sharing, and the CUs only need to carry out
the task of sharing the spectrum available to the designated AP with
other CUs. However, as we probably can speculate, no matter how
efficient such sharing scheme is, the overall system performance
might suffer because of the fixed and inefficient AP assignment. We
will see such performance degradation later in the simulation
section.

Nevertheless, we introduce two propositions stating two iterative
algorithms that enable the CUs to distributedly compute the NE power
allocation profile under the fixed $\mathbf{a}$. We refer the
readers to \cite{hong10m_japsa} for the proofs.
\newtheorem{P1}{Proposition}
\begin{P1}\label{propAIWF}
{\it For a fixed association profile $\mathbf{a}$, if in each
iteration $t$, the CUs in the set $\mathcal{N}_w$ iteratively do the
following.\\
1) Calculate the best reply power allocation:{\small
\begin{align}
\Phi^k_i({I}^t_{i}(k)) &\triangleq
\left[\frac{1}{\sigma_i}-\frac{n_w(k)+{I}_{i}^t(k)}{|h_{i,w}(k)|^2}\right]^+,
\forall~k\in\mathcal{K}_w\label{eqPlayerBestReplyAP}
\end{align}}
where $\sigma_i$ ensures
$\sum_{k\in\mathcal{K}_w}{\Phi^k_i({I}^t_{i}(k))}=\bar{p}_i$, and
let
$\bfPhi_i(\mathbf{I}^t_{i,w})\triangleq\left\{\Phi^k_i({I}^t_{i}(k))\right\}_{k\in\mathcal{K}_w}$.\\
2) Adjust their power profiles according to:{\small
\begin{align}
\mathbf{p}^{t+1}_{i,w}&=(1-\alpha_t)\mathbf{p}^{t}_{i,w}+\alpha_t\bfPhi_i(\mathbf{I}^t_{i,w})
\end{align}}
where the sequence $\{\alpha_t\}_{t=1}^{\infty}$ satisfy
$\alpha_t\in(0,1)$ and:{\small
\begin{align}
\hspace{-0.1cm}\lim_{t\to\infty}\alpha_t=0,~\lim_{T\to\infty}\sum_{t=1}^{T}\alpha_t=\infty,~
\lim_{T\to\infty}\sum_{t=1}^{T}\alpha^2_t<\infty.\label{eqAlpha}
\end{align}}
Then the CUs' individual power profiles converge to a NE power
allocation profile, i.e.,
$\lim_{t\to\infty}\mathbf{p}^t_{i,w}=\mathbf{p}^*_{i,w},~\forall~i\in\mathcal{N}_w$,
and
$\{\mathbf{p}^*_{i,w}\}_{i\in\mathcal{N}_w}\in\mathcal{E}_w(\mathbf{a})$.
We name the above algorithm the Averaged Iterative-Water Filling
(A-IWF).}
\end{P1}

\newtheorem{P2}{Proposition}
\begin{P1}\label{propSIWF}
{\it If in each iterative $t$, the CUs in the set $\mathcal{N}_w$
adjust their power profiles sequentially\footnote{By ``sequential"
we mean that the CUs in the set $\mathcal{N}_w$ take turns in
changing their power allocation, and only a single CU gets to act at
time $t$. All other CUs $j\ne i, j\in\mathcal{N}_w$ keep their power
allocation as in time $t-1$.} according to:
\begin{align}
\mathbf{p}^{t+1}_{i,w}&=\bfPhi_i(\mathbf{I}^t_{i,w}),
\end{align}
then their individual power profiles also converge to a NE power
allocation profile. We call the above algorithm the Sequential
Iterative-Water Filling (S-IWF).}
\end{P1}


From Proposition \ref{propAIWF} and \ref{propSIWF}, we conclude that
for a specific association profile $\mathbf{a}$, all CUs
$i\in\mathcal{N}$ are able to distributedly decide on their NE power
allocation profiles by running either the A-IWF or the S-IWF
algorithm. Several
comments regarding these algorithms are in order.\\
1) In order to calculate
$\left\{\Phi^k_i(.)\right\}_{k\in\mathcal{K}_w}$, in each iteration
individual CU only needs to know the aggregated {\it interference
plus noise} (IPN) contributed by all other CUs on the channels of
selected AP $w$,
$\left\{n_w(k)+{I}^t_{i}(k)\right\}_{k\in\mathcal{K}_w}$, and this
information can be fed back to the CUs $i\in\mathcal{N}_w$
by the AP $w$.\\
2) Consider a single AP $w$. We have shown in \cite{hong10m_japsa}
that when the number of channels becomes large, or equivalently the
portion of the spectrum belongs to this AP is very finely divided,
then the NE power allocation profile maximizes the sum capacity of
the AP. In another word, the NE is efficient. Similar observation
has been made in \cite{lai08}, where the authors proved that the NE
of a fading multiple-access water-filling game achieves capacity.
This somewhat surprising result, that selfish CUs by distributedly
allocate their power can achieve system capacity, provides
justification for the distributed spectrum sharing scheme analyzed
in this work.

\vspace{-0.1cm}
\subsection{The Joint AP Selection and Power Allocation
Algorithm}\label{subJASPA} We name the proposed algorithm Joint
Access Point Selection and Power Allocation (JASPA) algorithm.
Intuitively, the JASPA algorithm works as follows. For a fixed AP
association profile, all CUs calculate iteratively their NE power
allocations. After convergence, they individually try to see if they
can {\it strictly} increase their communication rates by switching
to another AP, {\it assuming that all other CUs keep their current
AP associations and power profiles}. When CU $i$ decides that its
next best AP association should be $w_i^*$, we record his decision
by a $W\times 1$ {\it best reply vector}
$\mathbf{b}_i:\mathbf{b}_i=\mathbf{e}_{w^*_i}$, where
$\mathbf{e}_{j}$ denotes a $W\times 1$ elementary vector with all
 entries $0$ except for the $j^{th}$ entry, which takes the value $1$. In the next iteration, CU
$i$'s AP association decision is made according to a $W\times 1$
{\it probability vector} $\bfbeta_i^t$, which is properly updated in
each iteration according to $\mathbf{b}_i$. We also suppose that
each CU has a length $M$ memory, operating in a first in first out
fashion, that records its last $M$ best reply vectors.

The proposed algorithm is detailed as follows.

1) {\bf Initialization}: Let t=0, CUs randomly choose their APs.

2) {\bf Calculation of the NE Power Allocation Profile}: Based on
the current association $\mathbf{a}^t$, all the CUs calculate their
NE power allocations $\mathbf{p}^{*}_{i}(\mathbf{a}^t)$, either by
A-IWF or S-IWF algorithm.

3) {\bf Selection of the Best Reply Association}: Each CU $i$ talks
to all the APs in the network, obtains necessary information in
order to find a set of APs $\mathcal{W}^t_i$ such that all
$w\in\mathcal{W}^t_i$ satisfies $w\ne\mathbf{a}^t(i)$ and:{\small
\begin{align}
\hspace{-0.5cm}\max_{\mathbf{p}_{i,w}\in\mathcal{F}_{i,w}}R_i(\mathbf{p}_{i,w},\mathbf{p}^*_{w}(\mathbf{a}^t);w)>
R_i(\mathbf{p}^*_{i}(\mathbf{a}^t),\mathbf{p}^*_{-i}(\mathbf{a}^t);{\mathbf{a}^t(i)}).\label{eqBetterAP}
\vspace{-0.5cm}
\end{align}
}
If $\mathcal{W}^t_i\ne\emptyset$, obtain the
$w^*_i\in\mathcal{W}^t_i$ that can offer the maximum rate (ties are
randomly broken); otherwise, let $w^*_i=\mathbf{a}^t(i)$. Set the
best choice vector $\mathbf{b}^{t+1}_i=\mathbf{e}_{w^*_i}$.

4) {\bf Update Probability Vector}: For each CU $i$, update the
$W\times 1$ probability vector $\bfbeta^t_i$ according to:{\small
\begin{align}
\bfbeta^{t+1}_i=\left\{ \begin{array}{ll} \bfbeta
^t_i+\frac{1}{M}(\mathbf{b}^{t+1}_i-
\mathbf{b}^{t-M}_i)&~\textrm{if}~M \le t\\
\bfbeta ^t_i+\frac{1}{M}(\mathbf{b}^{t+1}_i-
\mathbf{b}^{1}_i)&~\textrm{if}~M>t>0\\
\mathbf{b}^{1}_i &\textrm{if}~t=0. \\
\end{array} \right.\label{eqUpdateBeta}
\end{align}}
Shift $\mathbf{b}_i^{t+1}$ into the end of the memory; shift
$\mathbf{b}^{t-M}_i$ out from the front of the memory if $t\ge M$.

5) {\bf Determine the Next AP Association}: Each CU $i$ samples the
AP index for association at iteration $t+1$ based on
$\bfbeta^{t+1}_i$:{\small
\begin{align}
\mathbf{a}^{t+1}(i)\sim multi(\bfbeta^{t+1}_i),\label{eqSampleA}
\end{align}}
where $multi(.)$ represents a multinomial distribution.

6) {\bf Continue}: Let t=t+1, and go to Step 2).

%

We make the following remarks about the above algorithm.
\newtheorem{R1}{Remark}
\begin{R1}
It is crucial that each CU finally decides on choosing a single AP.
Failing to do so will result in system instability, in which the CUs
switch AP association indefinitely, and much of the system resource
will be wasted for closing old connections and establishing new
connections between the APs and CUs. Specifically, it is desirable
to have $ \lim_{t\to\infty}\bfbeta^t_i=\bfbeta^*_i,
\forall~i\in\mathcal{N}$, where $\bfbeta^*_i$ is an elementary
vector with a single entry $1$, and all other entries $0$.
\end{R1}

\newtheorem{R2}{Remark}
\begin{R1}
The best reply vectors $\mathbf{b}^{t+1}_i$ are decided in each
iteration based on the other CUs' AP associations and power profiles
in the previous iteration.  It can be straightforwardly shown that
in order to calculate
$\max_{\mathbf{p}_{i,w}\in\mathcal{F}_{i,w}}R_i(\mathbf{p}_{i,w},\mathbf{p}^*_{w}(\mathbf{a}^t);w)$
for different $w\in\mathcal{W}$, individual CU $i$ does not need to
know the strategies of all other CUs in the network, nor does it
need to know the system association profile $\mathbf{a}^t$. Instead,
it only requires the aggregated IPN on each channel from each AP of
the last iteration. This is precisely the necessary information
needed for finding the set $\mathcal{W}^t_i$ in Step 3) of the
JASPA. This property of the algorithm contributes to the reduction
of the amount of messages exchanged between APs and each CU when
making association decisions. \end{R1}
\newtheorem{R3}{Remark}
\begin{R1}
Considering the overhead regarding to end an old connection and
re-establish a new connection, it is reasonable to assume that a
selfish CU is unwilling to abandon its current AP if the new one
cannot offer {\it significant} improvement of the data rate. We can
model such unwillingness of the CUs by introducing a {\it connection
cost} $c_i\ge 0$, which is a private parameter for each CU $i$. A CU
$i$ will only seek to switch to a new AP if the new one can offer
rate improvement of at least $c_i$, i.e., it will only switch to
those APs $w\in\mathcal{W}^t_i$ that satisfies:{
\begin{align}
\hspace{-0.1cm}\max_{\mathbf{p}_{i,w}\in\mathcal{F}_{i,w}}\hspace{-0.2cm}
R_i(\mathbf{p}_{i,w},\mathbf{p}^*_{w}(\mathbf{a}^t);w)\ge
R_i(\mathbf{p}^*_{i}(\mathbf{a}^t),\mathbf{p}^*_{-i}(\mathbf{a}^t);{\mathbf{a}^t(i)})+c_i.\nonumber
\end{align}}
From a system point of view, such unwillingness to switch by the CUs
might contribute to improved convergence speed of the algorithm, but
might also result in reduced system throughput. These two
phenomenons are indeed observed in our simulations, please see
section \ref{secSimluation} for examples.
\end{R1}

\subsection{Global Convergence of the JASPA algorithm}\label{subsecConvergence}
In this subsection, we prove that our algorithm converges to a JEP {
globally}, i.e., the algorithm converges regardless of the initial
starting points or the realizations of the channel gains. Due to
space limit, we refer the readers to \cite{hong10m_japsa} for the
proofs of the Proposition \ref{propIO}.

We first introduce some notations.  Define a set $\mathcal{A}$ as
follows:{\small
\begin{align}
\mathbf{a}\in \mathcal{A} \Longleftrightarrow \mathbf{a}
\textrm{~~appears infinitely often in ~} \{\mathbf{a}^t\}.
\end{align}}
We state a proposition characterizing the set $\mathcal{A}$.
\newtheorem{P3}{Proposition}
\begin{P1}\label{propIO}
{\it Let $M\ge N$. Then at least one element in the set
$\mathcal{A}$, say $\mathbf{a}^{*}$, is a NE association profile.
Moreover, $(\mathbf{a}^{*},\mathbf{p}^{*}(\mathbf{a}^*))$ is a JEP
(satisfy equation \eqref{eqNE}).}
\end{P1}

Using the results in Proposition \ref{propIO}, we obtain the
following convergence results.
\newtheorem{T2}{Theorem}
\begin{T1}\label{theoremConvergenceJASPA}
{\it Let $M\ge N$. Then the JASPA algorithm produces a sequence
$\left\{(\mathbf{a}^t,
\mathbf{p}^*(\mathbf{a}^t))\right\}_{t=1}^{\infty}$ that converges
to a JEP $(\mathbf{a}^*,\mathbf{p}^*(\mathbf{a}^*))$ with
probability 1.}
\end{T1}
\begin{proof}
We first show that the sequence
$\left\{\mathbf{a}^t\right\}_{t=1}^{\infty}$ converges to an
equilibrium profile $\mathbf{a}^*$. Notice that if at time $T$,
$\mathbf{a}^{T}=\mathbf{a}^*$, and in the next $M$ iterations, we
always have $\mathbf{a}^{T+t}=\mathbf{a}^*$, then the algorithm
converges.

Let $\mathcal{A}^*\in\mathcal{A}$ contains all the NE association
profiles in $\mathcal{A}$. Let $\{\mathbf{a}^{t(k)}:k\ge 1\}$ be the
infinite subsequence satisfying $\mathbf{a}^{t(k)}\in\mathcal{A}^*$.
Without loss of generality, assume $t(k)-t(k-1)\ge M$. Let us denote
by $C_k$ the event in which the process converges to a
$\mathbf{a}^*\in\mathcal{A}^*$, after a sequence of best replies
equals to $\mathbf{a}^*$ of length $M$ occurs, starting at time
$t(k)$: $ C_k=\bigcap_{l=1}^{M}\{\mathbf{a}^{t(k)+l}=\mathbf{a}^*\}.
$ Note, $ Pr(C_{k+1}|C_k^c)\ge(\frac{1}{M})^{N\times M}$, because
whenever $\mathbf{a}^*$ appears, each CU $i$'s best reply should be
$\mathbf{a}^*(i)$, hence $\mathbf{a}^*(i)$ will be inserted into the
last slot of CU $i$'s memory. Then with probability
$(\frac{1}{M})^N$, all CUs sample the last memory and $\mathbf{a}^*$
will appear in the next iteration. Thus,{\small
\begin{align}
\hspace{-0.1cm}&Pr\hspace{-0.1cm}\left(\bigcap_{k\ge
1}C^c_k\right)\hspace{-0.15cm}=\hspace{-0.15cm}\lim_{T\to\infty}\hspace{-0.15cm}Pr\hspace{-0.1cm}\left(\bigcap_{k=1}^{
T}C^c_k\right)\hspace{-0.1cm}=\hspace{-0.1cm}\lim_{T\to\infty}\prod_{k=1}^{T-1}\hspace{-0.15cm}\left(1-Pr(C_{k+1}|C^c_k)\right)\nonumber\\
\hspace{-0.1cm}&\le\lim_{T\to\infty}\Big(1-(\frac{1}{M})^{n\times
M}\Big)^{T-1}\hspace{-0.3cm}=0.
\end{align}}
This says $ Pr(\mathbf{a}^t ~\textrm{converges to a }
\mathbf{a}^*\in\mathcal{A}^*~\textrm{eventually})=1. $ Finally,
because $\mathbf{p}^*(\mathbf{a}^*)\in\mathcal{E}(\mathbf{a}^*)$ is
a NE power allocation profile, we conclude that
$\left(\mathbf{a}^*,\mathbf{p}^*(\mathbf{a}^*)\right)$ is a JEP.
\end{proof}
%

Now that we have shown the convergence of the JASAP to the JEP, it
is of interest to evaluate the ``quality" of such equilibrium. In
this work, we use the system throughput to measure the quality of
the JEP, and our simulation results (to be shown in section
\ref{secSimluation}) are very encouraging.

\section{Extensions to the JASPA Algorithm}\label{secExtension}
The JASPA algorithm presented in the previous section is
``distributed" in the sense that the computation that each CU needs
to carry out in each iteration only requires some local/summary
information, i.e., the aggregated IPN at different APs in different
channels, and the CU's own channel gain. However, this algorithm
requires that for each AP association profile $\mathbf{a}^t$, an
{\it intermediate} equilibrium $\mathbf{p}^*(\mathbf{a}^t)$ should
be reached (in Step 2), and  the CUs {\it cannot} choose their next
AP association profile {\it until} the system reaches such
equilibrium. This requirement poses a relatively strong level of
{\it coordination} among the CUs, which is not very desirable for a
distributed algorithm.

In this section, we propose the following two algorithms that do not
require the CUs reach any intermediate equilibria: 1) a sequential
version of the JASPA algorithm (Se-JASPA) in which CUs act one by
one in each iteration, and 2) a simultaneous/parallel version of the
JASPA algorithm (Si-JASPA) in which CUs act at the same time.

The Se-JASPA algorithm is detailed in Table \ref{tableSeJASPA}.
\begin{table}[ht]
\begin{center}
\vspace{-0.2cm}
\begin{tabular*}{ 0.5\textwidth }{l}
\hline\\
1) {\bf Initialization (t=0)}: Each CU randomly chooses
$\mathbf{a}^0(i)$ and
$\mathbf{p}_{i,\mathbf{a}^0(i)}^0$\\
2) {\bf Determine the Next AP Association}:\\
~~If it is CU $i$'s turn to act, (e.g., $\{(t+1) \textrm{mode} N
\}+1=i $), then CU $i$ \\
~~finds a set $ \mathcal{W}^*_i$ s.t.:\\
~~~~~~~~$ \mathcal{W}^*_i=
\arg\max_{w\in\mathcal{W}}\max_{\mathbf{p}_{i,w}\in\mathcal{F}_{i,w}}
R(\mathbf{p}_{i,w},\mathbf{p}^t_{w};w)\label{eqUpdateASequential}
$\\
~~It selects an AP by randomly picking
$w^*\in\mathcal{W}^*_i$ and setting $\mathbf{a}^{t+1}(i)=w^*$.\\
~~For other CUs $j\ne i$, $\mathbf{a}^{t+1}(j)=\mathbf{a}^{t}(j)$\\
3) {\bf Update the Power Allocation}: \\
~~Denote $w^*=\mathbf{a}^{t+1}(i)$, Then CU $i$ calculates $\mathbf{p}^{t+1}_i$ as\\
~~~~$ \mathbf{p}^{t+1}_i=\left\{ \begin{array}{l} \arg
\max_{\mathbf{p}_{i,w^*}\in\mathcal{F}_{i,w^*}}R_i(\mathbf{p}_{i,w^*},\mathbf{p}^t_{w^*};{w^*}),~
\textrm{if}~ w^*\ne \mathbf{a}^{t}(i) \\
\arg
\max_{\mathbf{p}_{i,w^*}\in\mathcal{F}_{i,w^*}}R_i(\mathbf{p}_{i,w^*},\mathbf{p}^t_{-i,w^*};{w^*}),
~\textrm{otherwise}\\
\end{array} \right.\label{eqUpdatePSequential}
$ \\
~~For other CUs $j\ne i$, $\mathbf{p}_j^{t+1}=\mathbf{p}_j^t$\\

4) {\bf Continue}: Let t=t+1, and go to Step 2)\\

 \hline
\end{tabular*}
\caption{The Se-JASPA Algorithm} \label{tableSeJASPA}
\end{center}
\vspace{-0.9cm}
\end{table}

We partially characterize the convergence behavior of Se-JASPA
algorithm in the following theorem, the proof of which can be found
in \cite{hong10m_japsa}.

\newtheorem{T3}{Theorem}
\begin{T1}\label{theoremConvergenceSeJASPA}
{\it The sequence $\{P(\mathbf{p}^t,\mathbf{a}^t)\}^{\infty}_{t=1}$
produced by the Se-JASPA algorithm is non-decreasing and
converging.}
\end{T1}

We see that the Se-JASPA algorithm differs from the JASPA algorithm
in several important ways. Firstly, a CU $i$ does not need to keep
its best reply vector $\mathbf{b}^t$ as it does in JASPA. It decides
on its AP association greedily in step 2). Secondly, a CU $i$, after
deciding a new AP $\mathbf{a}^{t+1}(i)=w^*$, does not need to go
through the process of reaching an intermediate equilibrium with all
other CUs to obtain $\mathbf{p}^{*}_{i}(\mathbf{a}^{t+1})$. However,
the CUs still need to be coordinated for the exact sequence of their
update, because in each iteration only a single CU is allowed to
act. Such order of update can be agreed upon and enforced by the APs
in the network. As might be inferred by the sequential nature of
this algorithm, when the number of CUs is large, the convergence
time becomes long.

The Si-JASPA algorithm, as detailed in Table \ref{tableSiJASPA},
overcomes the above difficulties arise in Se-JASPA. We note that in
the algorithm, the variable $T_i$ represents the duration that CU
$i$ has stayed in the current AP, and the stepsize $\alpha_t$
satisfies \eqref{eqAlpha}.

\begin{table}
\begin{center}
\vspace{-0.1cm}
\begin{tabular*}{ 0.5\textwidth }{l}
\hline\\
1) {\bf Initialization (t=0)}: Each CU $i$ randomly chooses
$\mathbf{a}^0(i)$ and $\mathbf{p}_{i,\mathbf{a}^0(i)}^0$
\\
2) {\bf Selection of the Best Reply Association}:\\
~~Each CU obtains the AP $w^*_i$ and set $\mathbf{b}^{t+1}_i$ following Step 3) of JASPA\\
3) {\bf Update Probability Vector}: \\
~~Each CU $i$ updates the probability vector $\bfbeta^t_i$ according
to \eqref{eqUpdateBeta} \\
~~Shift $\mathbf{b}_i^{t+1}$ into the memory; shift
$\mathbf{b}^{t-M}_i$ out of memory if $t\ge M$
\\
4) {\bf Determine the Next AP Association}: \\
~~Each CU $i$ samples the AP index for association as in
\eqref{eqSampleA}
\\
5) {\bf Compute the Best Reply Power Allocation}: \\
~~Let $w^{t+1}_i=\mathbf{a}^{t+1}(i)$. Each CUs $i$ calculates
$\mathbf{p}^*_i$ as\\
~~~~~$
\mathbf{p}^*_i=\max_{\mathbf{p}_{i,w^{t+1}_i}}R_i(\mathbf{p}_{i,w^{t+1}_i},\mathbf{p}^t_{-i,w^{t+1}_i};{w^{t+1}_i})
$
\\
6) {\bf Update the Duration of Stay}: \\
~~Each CU $i$ maintains and updates a variable $T_i$: \\
~~~~${T}_i=\left\{
\begin{array}{ll}
1&\textrm{if}~\mathbf{a}^{t+1}(i)\ne\mathbf{a}^{t}(i)\\
T_i+1&\textrm{if}~\mathbf{a}^{t+1}(i)=\mathbf{a}^{t}(i)\\
\end{array} \right.
$
\\
7) {\bf Update the Power Allocation}: \\
~~Each CU $i$ calculates $\mathbf{p}^{t+1}_i$ as follows:
\\
~~~~$\mathbf{p}^{t+1}_i=\left\{
\begin{array}{ll}
\mathbf{p}^*_i&~\textrm{if}~\mathbf{a}^{t+1}(i)\ne\mathbf{a}^{t}(i)\\
(1-\alpha_{T_i})\mathbf{p}^t_i+\alpha_{T_i}\mathbf{p}^*_i&~~\textrm{if}~\mathbf{a}^{t+1}(i)=\mathbf{a}^{t}(i)\\
\end{array} \right.\label{eqUpdatePSimultaneous}
$\\
8) {\bf Continue}: Let t=t+1, and go to Step 2)\\
 \hline
\end{tabular*}
\caption{The Si-JASPA Algorithm} \label{tableSiJASPA}
\end{center}
\vspace{-1.3cm}
\end{table}
The structure of the Si-JASPA is almost the same as the JASPA except
that each CU, after switching to a new AP, does not need to go
through the process of joint computation of the intermediate
equilibrium solution with all other CUs currently associated with
the same AP: they can choose their AP ``continuously". The level of
coordination among the CUs required for this algorithm is minimum
among all the three algorithms. The simultaneous update required by
this algorithm can be realized by either one of the following
approaches:\\
1) The APs agree upon the update interval off-line. Each CU is
equipped with a timer. The first time a CU comes into the system, it
is informed by its initial associated AP the update interval and the
next update instance. Then the CU can perform update on its
own.\\
2) The APs agree upon the update interval off-line, and they alert
the CUs associated with them the update instances by broadcasting.

Extensive simulations confirm that generally this algorithm
converges faster than the Se-JASPA.

\vspace{-0.1cm}
\section{Simulation Results}\label{secSimluation}
In this section, we present simulation results to validate the
proposed algorithms. For each experiment we show the results
obtained by running either Si-JASPA and Se-JASPA, or by the original
JASPA.

We have the following general settings for the simulation. We place
multiple CUs and APs randomly in a $10m\times10m$ area; we let
$d_{i,w}$ denote the distance between CU $i$ and AP $w$, then the
channel gains between CU $i$ and AP $w$,
$\{|h_{i,w}(k)|^2\}_{k\in\mathcal{K}_w}$, are independently drawn
from an exponential distribution with mean $\frac{1}{d^2_{i,w}}$.
We let the available channels to be evenly pre-assigned to different
APs. We let the length of the individual memory to be $M=10$. For
ease of presentation and comparison, when we use the JASPA algorithm
with connection cost, we set all the CUs' connection cost
$\{c_i\}_{i\in\mathcal{N}}$ to be identical. In the following, when
we say a ``snapshot" of the network, we refer to the network with
fixed (but randomly generated as above) AP, CU locations and channel
gains.

\begin{figure}[htb]
\vspace{-0.2cm} {\includegraphics[width=
0.7\linewidth]{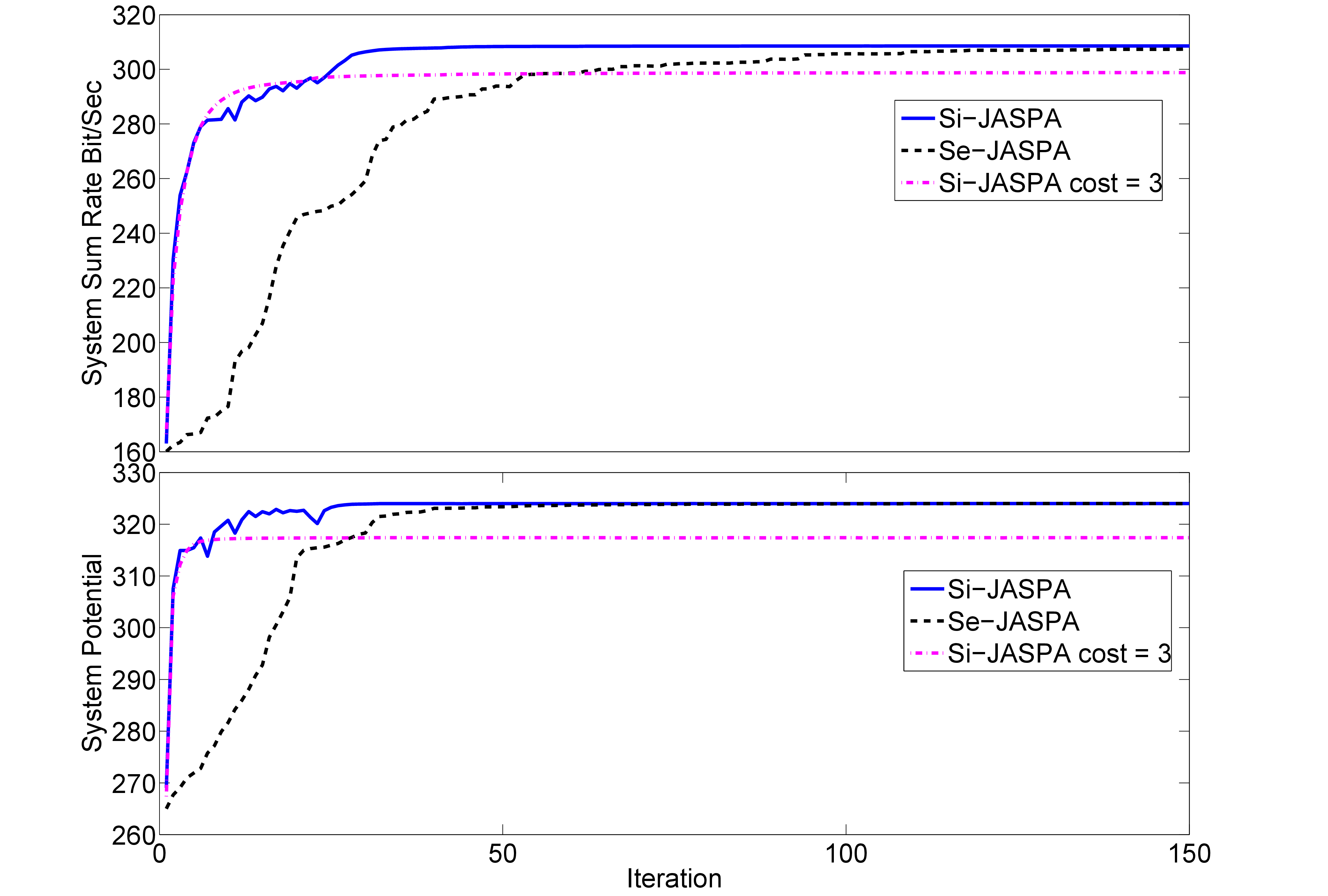}}
\vspace*{-.3cm}\caption{Convergence of different algorithms. Top:
evolution of system sum rate. Bottom: evolution of the value of
system potential function. }\label{figCompareSpeed} \vspace*{-.5cm}
\end{figure}

We first show the results regarding to the convergence of the
algorithm. We only show the results for Si-JASPA and Se-JASPA in
this experiment. We first consider a network with $20$ CUs, $64$
channels, and $4$ APs. Fig. \ref{figCompareSpeed} shows the
evolution of the system throughput as well as the values of the
system potential function generated by a typical run of the
Se-JASPA, Si-JASPA and Si-JASPA with connection cost $c_i=3$
bit/sec. We observe that the two Si-JASPA based algorithms converge
very fast, while the Se-JASPA converges very slowly. After
convergence, the system throughput achieved by Si-JASPA with
connection cost is smaller than that of the other two algorithms.
Notice that in the bottom part of Fig. \ref{figCompareSpeed}, the
system potential generated by the Se-JASPA is non-decreasing with
respect to the iterations. This phenomenon has been predicted in
Theorem \ref{theoremConvergenceSeJASPA}.

\begin{figure}[htb]
\vspace{-0.3cm} {\includegraphics[width=
0.7\linewidth]{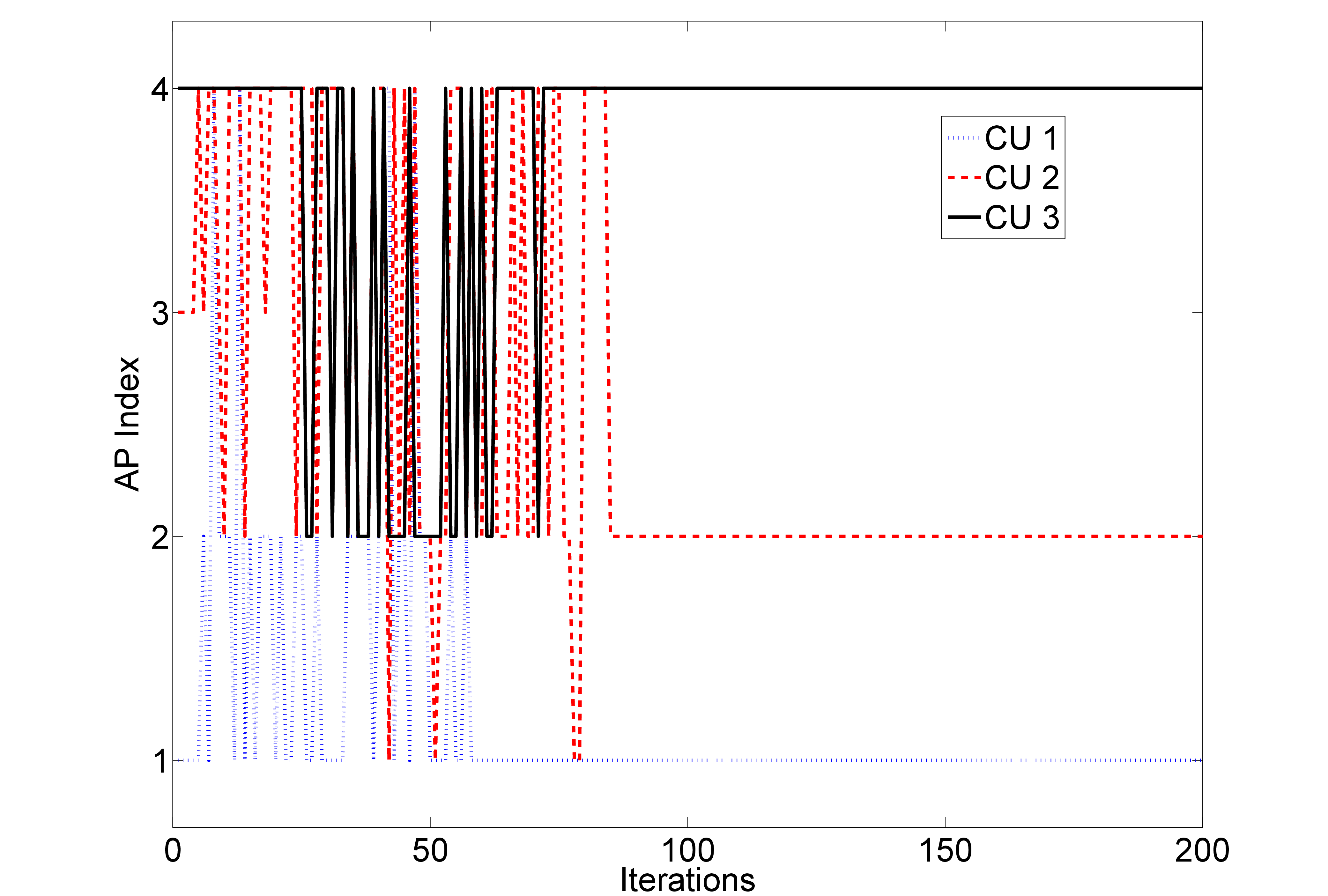}}
\vspace*{-.3cm}\caption{Convergence of Selected CUs' AP
selection.}\label{figConvergenceSelection} \vspace*{-.2cm}
\end{figure}
%

\begin{figure*}[htb] \vspace*{-.2cm}
        \begin{minipage}[t]{0.3\linewidth}
    \centering
    {\includegraphics[width=
1\linewidth]{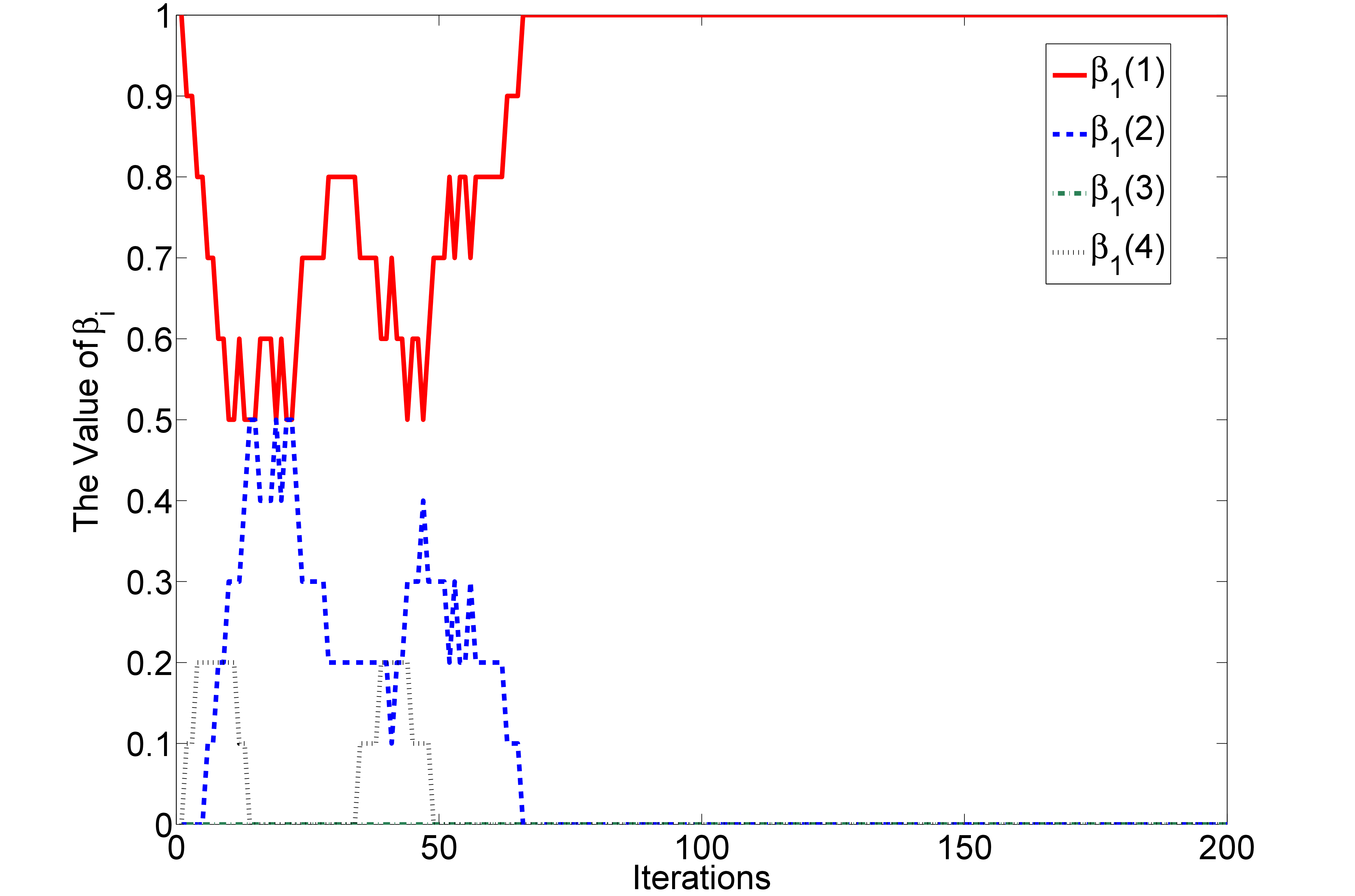}
\vspace*{-.5cm}}
\end{minipage}
\begin{minipage}[t]{0.3\linewidth}
    \centering
    {\includegraphics[width=
1\linewidth]{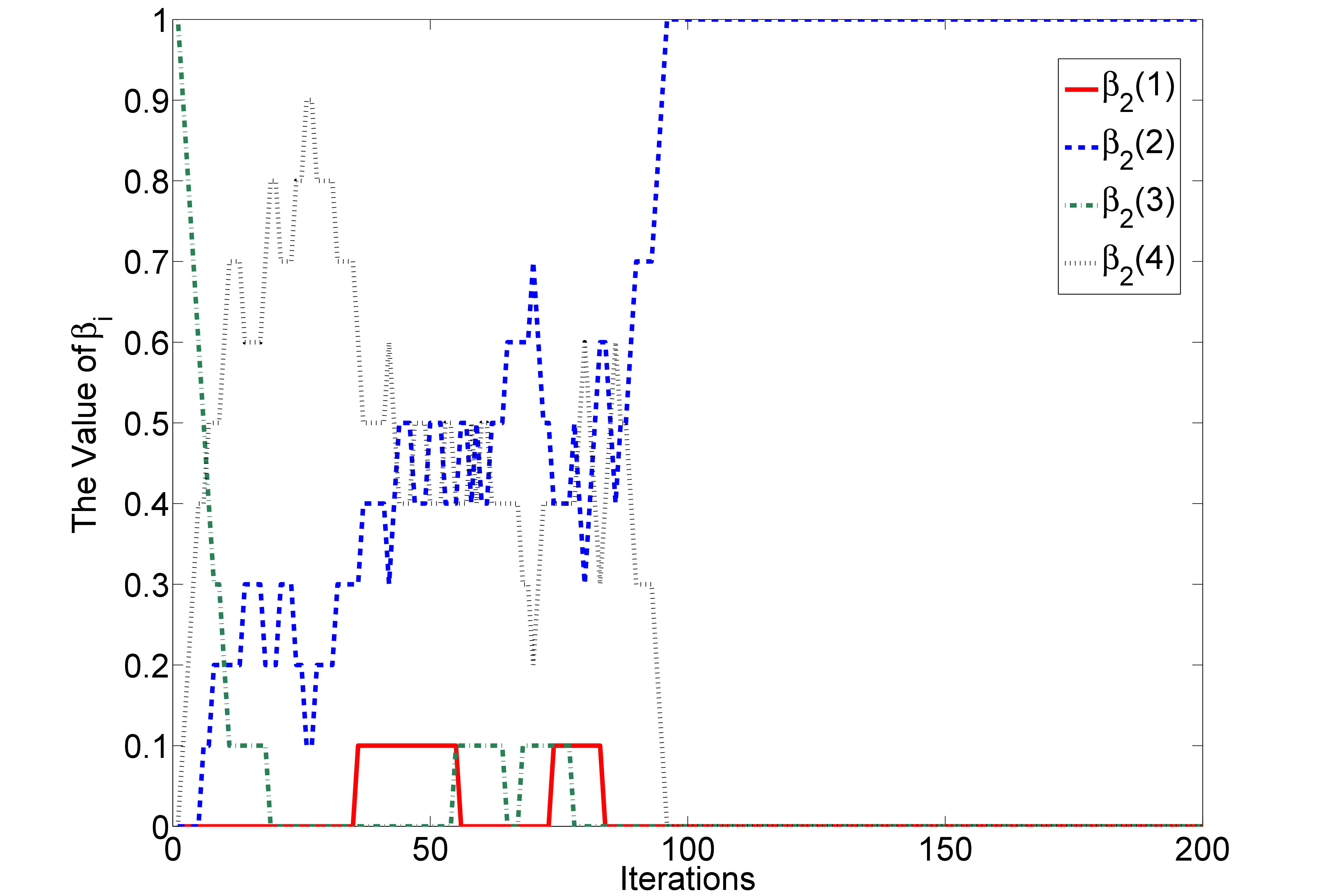}
\vspace*{-.5cm}}
\end{minipage}
    \begin{minipage}[t]{0.3\linewidth}
    \centering
    {\includegraphics[width=
1\linewidth]{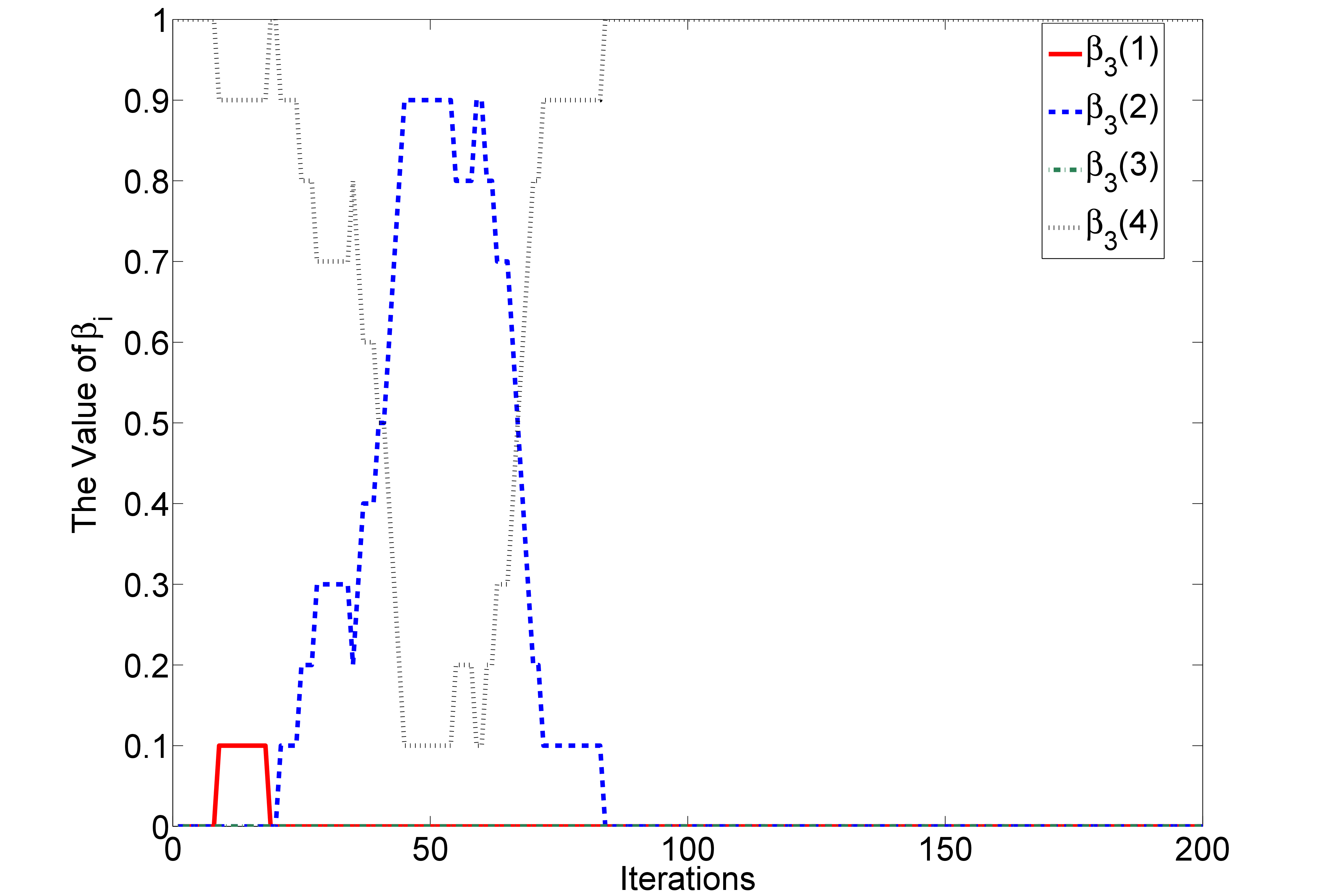}
\vspace*{-0.5cm}}
\end{minipage}
\vspace*{-0.2cm}\caption{Convergence of the probability vector
$\beta^t_1$, $\beta^t_2$ and $\beta^t_3$ of CUs' 1, 2,
3.}\label{figConvergenceBeta}\vspace*{-0.6cm}
    \end{figure*}

\begin{figure}[!t]
\vspace{-0.2cm} {\includegraphics[width=
0.7\linewidth]{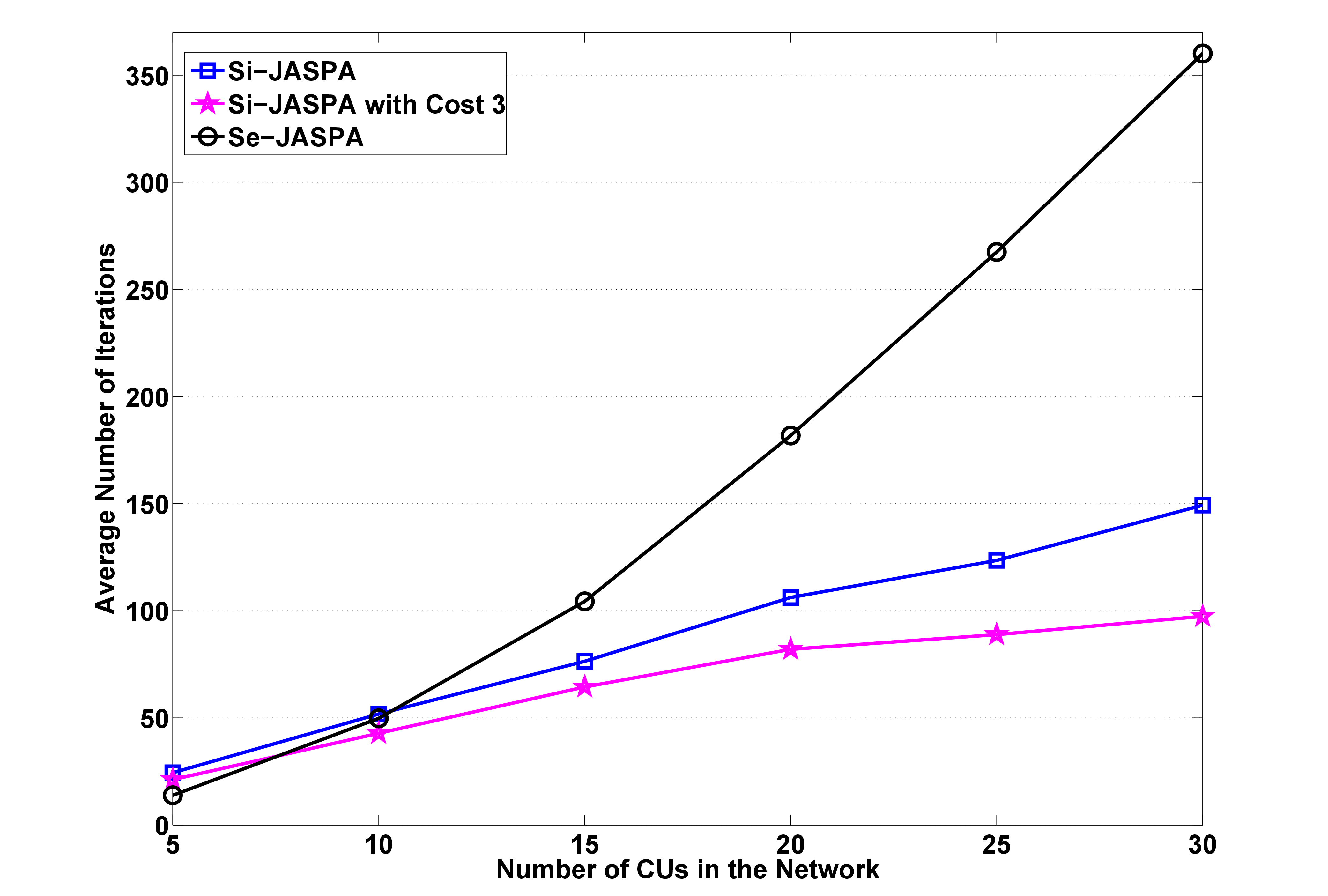}}
\vspace*{-.3cm}\caption{Comparison of averaged convergence
speeds.}\label{figConvergenceSpeed} \vspace*{-.4cm}
\end{figure}

Fig. \ref{figConvergenceSelection} shows the evolution of the AP
selections made by the CUs in the network during a typical run of
the Si-JASPA algorithm. We only show 3 out of 20 CUs (we call the
selected CUs CU 1, 2, 3 for easy reference) in order not to make the
figure overly crowded. Fig. \ref{figConvergenceBeta} shows the
corresponding evolution of the probability vectors
$\{\bfbeta^t_i\}_{t=1}^{100}$ for the three of the CUs selected in
Fig. \ref{figConvergenceSelection}. It is clear that upon
convergence, all the probability vector converges to an elementary
vector.
We then evaluate how the number of CUs in the network affects the
speed of convergence of different algorithms. In order to do so, we
compare the average iterations to achieve convergence in the network
with 4 APs, 64 channels and different number of CUs, for the
following three algorithms 1) Si-JASPA, 2) Se-JASPA, 3) Si-JASPA
with connection cost $c_i=3$ bit/sec for all CUs. From
Fig.\ref{figConvergenceSpeed}, we see  that when the number of CUs
in the system becomes large, the sequential version of the JASPA
takes significantly longer time to converge than the two
simultaneous versions of the JASPA algorithm. We can also see that
the connection costs adopted by individual CUs indeed have positive
effects on the convergence speed of the system.
Note each point in this figure represents the average of 100
independent runs of each algorithm on randomly generated network
snapshots.

We subsequently evaluate the network throughput performance
achievable by the JEP computed by the JASPA.

\begin{figure}[!t]
\vspace{-0.0cm}{\includegraphics[width=
0.7\linewidth]{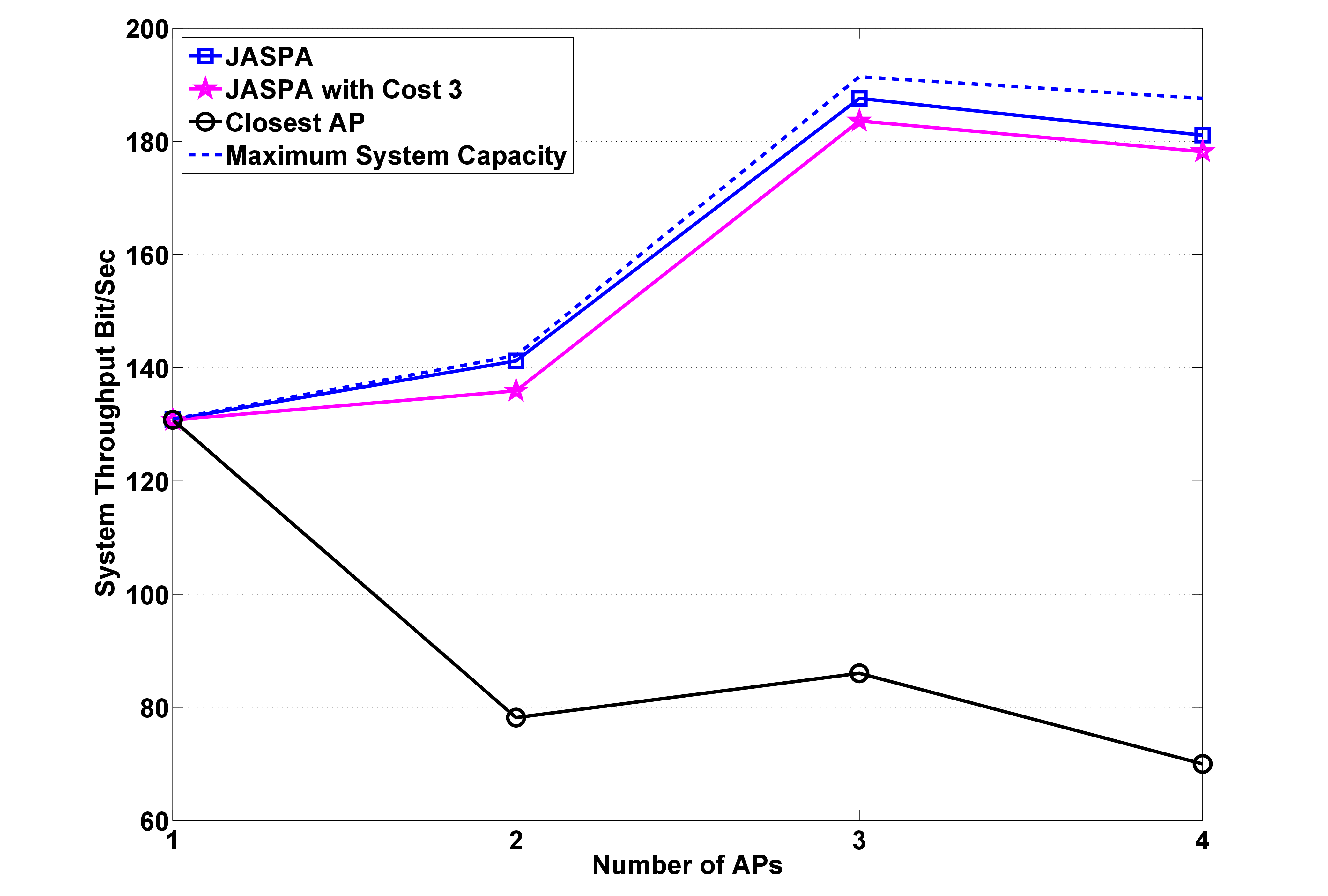}}
\vspace*{-.3cm}\caption{Comparison of averaged system throughput by
different algorithms with the throughput upper bound in a 8 CU
network.}\label{figCompareThrouputSmallScale} \vspace*{-.5cm}
\end{figure}

\begin{figure}[!t]
{\includegraphics[width=
0.8\linewidth]{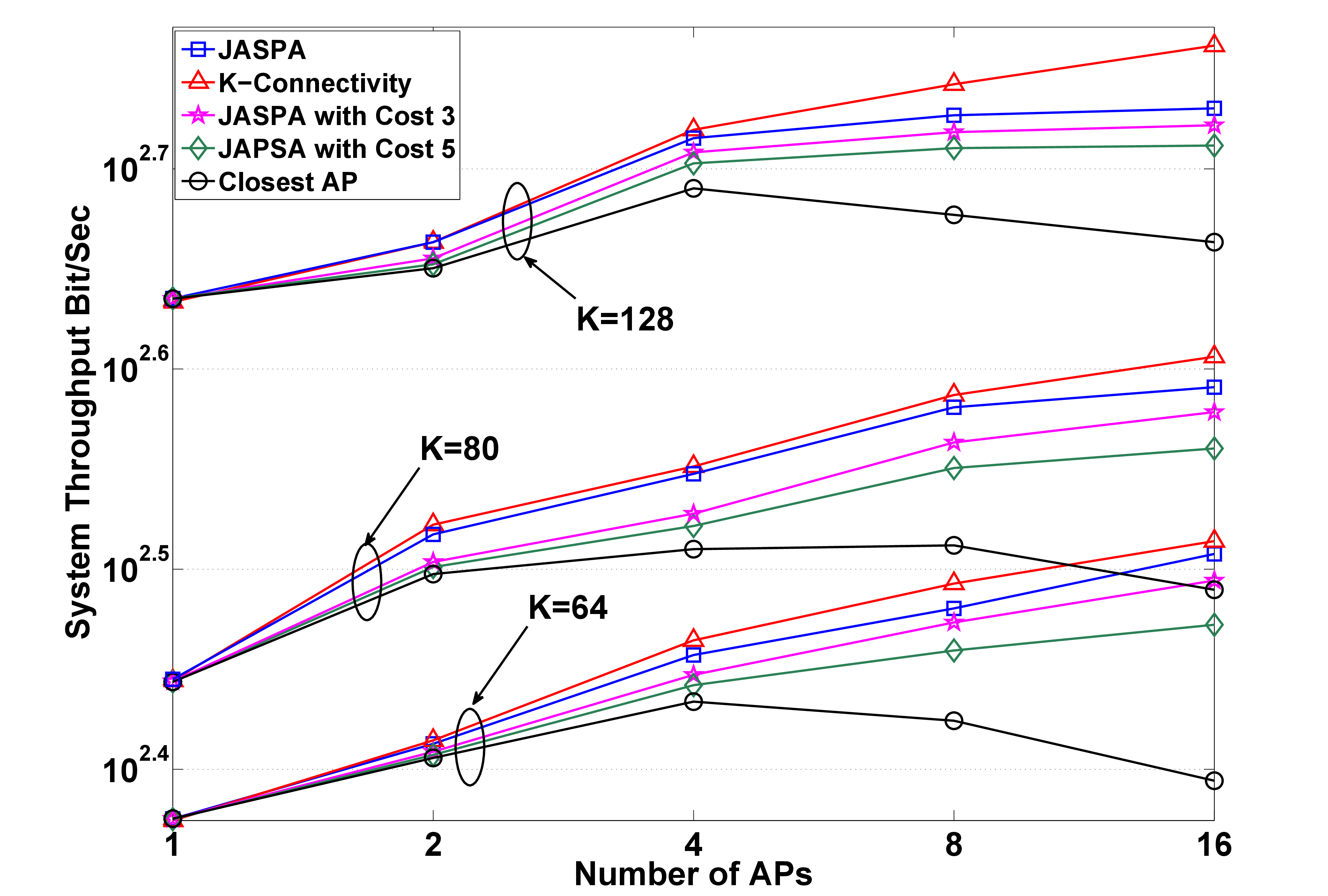}}
\vspace*{-.3cm}\caption{Comparison of the system throughput
\newline in a 30 CU network.}\label{figCompareThrouput} \vspace*{-.6cm}
\end{figure}

We first investigate a small networks with $8$ CUs, $64$ channels
and $1,~2,~3,~4$ APs, and compare the performance of JASPA related
algorithms to the maximum network throughput that can be achieve for
the same network. The maximum network throughput for a snapshot of
the network is calculated by the following two steps: 1) for a
specific AP-CU association profile, say $\mathbf{a}$, calculate the
maximum network throughput (denoted by $T(\mathbf{a})$) by summing
up the maximum capacity\footnote{For a single AP with fixed number
of users and channel gains, the maximum capacity is the well-known
multiple access channel sum capacity.} of individual APs in the
network; 2) enumerate {\it all possible} AP-CU association profiles,
and find $T^*=\max_{\mathbf{a}}T(\mathbf{a})$. We see the reason
that we choose to focus on such relatively small networks in this
experiment is that for a large network, the time it takes for the
above exhaustive search procedure to find the maximum network
throughput becomes prohibitive. The result is shown in
Fig.\ref{figCompareThrouputSmallScale}, where each point in the
figure is obtained by averaging the results obtained by the
algorithms on 100 independent snapshots of the network. We see that
the JASPA algorithm performs very well with little throughput loss,
while the closest AP algorithm, which separates the tasks of
spectrum decision and spectrum sharing, performs poorly.
We then start to look at the performance of larger networks with 30
CUs, up to 16 APs and up to 128 channels. Fig.
\ref{figCompareThrouput} shows the comparison of the performance of
JASPA, JASPA with individual cost $c_i=3$ bit/sec and $c_i=5$
bit/sec, and the closest AP algorithm. We adopt the actual distance
as the measure of ``closeness" in the closest AP algorithm. Each
point in this figure is the average of 100 independent runs of the
algorithms.
Due to the prohibitive computation time required, we are unable to
obtain the maximum system throughput for these relatively large
networks. We instead compute the equilibrium system throughput that
can be achieved in a non-cooperative game {\it if all CUs are able
to connect to multiple APs at the same time}. We refer to this
situation the {\it K-connectivity} network (the ``K-Connectivity"
network is similar in spirit to the ``multi-homing" WLAN studied in
\cite{shakkottai07}). It is clear that in the K-connectivity case,
there is no need for the CUs to perform the AP selection algorithm,
and the CUs in this network enjoy the flexibility of being able to
connect to multiple APs at the same time. However, we observe that
the performance of JASPA is very close to that of the
``K-Connectivity" network.
%

From Fig. \ref{figCompareThrouput} we see that when the number of
APs increases, the throughput of the JASPA algorithm becomes much
better than the closest AP algorithm, a phenomenon that is partly
due to the fact that for the closest AP algorithm, the separation of
the AP selection and power allocation process results in the
insufficient use of the spectrum: when the number of AP increases,
it becomes increasingly more probable that several APs are idle
because no CUs are close to them. Fig. \ref{figCompareThrouput},
along with Fig. \ref{figConvergenceSpeed}, also serve to confirm our
early speculation that there indeed exists tradeoff of convergence
speed and system throughput between JASPA and JASPA with connection
cost.


\vspace{-0.05cm}
\section{Conclusion}\label{secConclusion}
In this paper, we addressed the joint AP association and power
allocation problem in a CRN, and formulate it into a non-cooperative
game with hybrid strategy space. We characterized the NE of this
game, and provided distributed algorithms to reach such equilibrium.
Empirical evidence gathered from simulation experiments suggests
that the equilibrium has very promising quality in term of the
system throughput.

The problem analyzed in this work, particularly the game with mixed
strategy space, can be extended to solve many other problems, for
example, the CRN with interference channel and segmented spectrum
mentioned in section \ref{subMotivation}. It will also be our future
research topic to analyze the effect of random arrivals and
departures of the CUs on the performance of the algorithm, and
propose suitable heuristic dealing with these situations.

\bibliographystyle{IEEEtran}
\bibliography{ref}

\end{document}